\documentclass{elsart}
\usepackage{graphics,color,epsfig,amssymb}
\newcounter{bla}

\def\be{\begin{equation}}
\def\ee{\end{equation}}
\def\bea{\begin{eqnarray}}
\def\eea{\end{eqnarray}}
\def\nn{\nonumber}

\newcommand{\idel}{i\,\delta}
\newcommand{\eps}{\epsilon}
\newcommand{\rd}{{\mathrm{d}}}

\begin{document}

\begin{frontmatter}
\hfill{IPPP/10/94, DCPT/10/188}\\ 

\title{SecDec: A general program for sector decomposition}

\author[a]{Jonathon Carter},
\author[b]{Gudrun Heinrich}

\address[a]{IPPP, Department of Physics, University of Durham, Durham DH1 3LE, UK}
\address[b]{Max-Planck-Institute for Physics, F\"ohringer Ring 6, 80805 Munich, Germany}

\begin{abstract}

We present a program for the numerical evaluation of multi-dimensional polynomial parameter integrals. 
Singularities regulated by dimensional regularisation are extracted using 
iterated sector decomposition.  The program evaluates the coefficients of 
a Laurent series in the regularisation parameter. It can be applied to multi-loop 
integrals in Euclidean space as well as other parametric integrals, e.g. phase space integrals.

\begin{flushleft}
PACS: 12.38.Bx, 
02.60.Jh, 	
02.70.Wz 	
\end{flushleft}

\begin{keyword}
Perturbation theory, Feynman diagrams, infrared singularities, multi-dimensional parameter integrals, numerical integration
\end{keyword}

\end{abstract}

\end{frontmatter}

{\bf PROGRAM SUMMARY}

\begin{small}
\noindent
{\em Manuscript Title: SecDec: A general program for Sector Decomposition}                                       \\
{\em Authors: J.~Carter, G.~Heinrich}                                                \\
{\em Program Title: SecDec}                                          \\
{\em Journal Reference:}                                      \\
{\em Catalogue identifier:}   AEIR\_v1\_0                                \\
{\em Licensing provisions: Standard CPC licence, http://cpc.cs.qub.ac.uk/licence/licence.html}  \\
{\em Programming language: Wolfram Mathematica, perl, Fortran}        \\
{\em Computer: from a single PC to a cluster, depending on the problem}  \\
{\em Operating system: Unix, Linux}                                       \\
{\em RAM:} depending on the complexity of the problem                                              \\
{\em Keywords:}  Perturbation theory, Feynman diagrams, infrared singularities, multi-dimensional parameter integrals, numerical integration
 \\
{\em PACS:}      
12.38.Bx, 
02.60.Jh, 	
02.70.Wz \\
{\em Classification:}                                         \\
  4.4 Feynman diagrams, 
  5 Computer Algebra, 
  11.1 General, High Energy Physics and Computing.\\

{\em Nature of problem:}\\
  Extraction of ultraviolet and infrared singularities from parametric integrals 
  appearing in higher order perturbative calculations in gauge theories, e.g. 
  multi-loop Feynman integrals, Wilson loops, phase space integrals.\\
   \\
{\em Solution method:}\\
 Algebraic extraction of singularities in dimensional regularisation using iterated sector decomposition. 
 This leads to a Laurent series in the dimensional regularisation parameter $\epsilon$, 
 where the coefficients are finite integrals over the unit-hypercube. 
 Those integrals are evaluated numerically by Monte Carlo integration.
   \\
{\em Restrictions:} Depending on the complexity of the problem, limited by 
memory and CPU time. Multi-scale integrals can only be evaluated at Euclidean points.\\
   \\
{\em Running time:}\\
Between a few minutes and several days, depending on the complexity of the problem.\\
   \\
\end{small}

\newpage

\hspace{1pc}
{\bf LONG WRITE-UP}

\section{Introduction}

Sector decomposition is an algorithmic method to isolate 
divergences from  parameter integrals as they occur for instance  
in perturbative quantum field theory.  
Originally it was devised by Hepp~\cite{Hepp:1966eg} in the context of the 
the proof of the BPHZ theorem in order to 
disentangle overlapping ultraviolet singularities.
Similar ideas, applied to the subtraction of infrared divergences, can be found 
e.g. in \cite{Speer:1977uf}. 
It was employed later to
extract logarithmic mass singularities from massive multi-scale integrals 
in the high energy limit at two loops~\cite{Roth:1996pd,Denner:2004iz}.

In~\cite{Binoth:2000ps}, the concept of sector decomposition 
was elaborated 
to a general algorithm in the context of dimensional regularisation, 
allowing  the isolation of ultraviolet as well as infrared singularities 
from Feynman parameter integrals in an automated way. 
First applications of this algorithm were the numerical evaluation of  
two-loop  box diagrams at certain Euclidean points,
see e.g. \cite{Binoth:2000ps,Binoth:2003ak,Heinrich:2004iq}.  
More recently, the method has been used to numerically check a number 
of analytic three-loop and four-loop results\,\cite{Gehrmann:2006wg,Heinrich:2007at,Heinrich:2009be,Baikov:2009bg,Czakon:2006pa,Boughezal:2006xk,Asatrian:2006ph,Smirnov:2010hd,Baikov:2010hf,Lee:2010cga,Lee:2010ug,Lee:2010ik,Gehrmann:2010ue,Gehrmann:2010tu},  
most of them produced by 
either the public program {\small FIESTA}\,\cite{Smirnov:2008py,Smirnov:2009pb}
or the code which is described in the present article.
Further references about recent applications of sector decomposition to multi-loop calculations 
can be found in \cite{Smirnov:2009pb,Heinrich:2008si}.

Sector decomposition also has been combined with other methods for a 
numerical calculation 
of loop  amplitudes, first on a diagrammatic level in
 Refs.~\cite{Ferroglia:2002mz,Binoth:2002xh}, 
later for whole amplitudes in  Refs.~\cite{Lazopoulos:2008de,Lazopoulos:2007ix,Anastasiou:2007qb,Anastasiou:2008rm}. 
The latter approaches contain a combination of sector decomposition and contour 
deformation~\cite{Soper:1998ye,Soper:1999xk,Binoth:2005ff,Nagy:2006xy,Gong:2008ww}, 
which allows one to integrate the Feynman parameter representation of an amplitude 
numerically in the physical region.  

As phase space integrals in $D$ dimensions can be written 
as dimensionally regularised 
parameter integrals, sector decomposition can also serve to factorise entangled 
singularity structures in the case of soft and collinear real radiation. 
This idea was first presented in~\cite{Heinrich:2002rc} and was subsequently 
applied to 
calculate all master  four-particle phase space integrals where up to two 
particles in the final state
can become soft and/or collinear~\cite{GehrmannDeRidder:2003bm}.
Shortly after, this approach has been extended to be applicable to 
exclusive final states as well 
by expressing the functions produced by sector decomposition 
in terms of distributions~\cite{Anastasiou:2003gr}. 
Further elaboration on this 
approach~\cite{Binoth:2004jv,Anastasiou:2004qd}  has 
lead to differential NNLO results for a number of processes~\cite{Anastasiou:2004xq,Anastasiou:2005qj,Anastasiou:2005pn,Anastasiou:2007mz,Melnikov:2006di,Melnikov:2006kv,Melnikov:2008qs,Biswas:2009rb,Gavin:2010az}.
The combination of the Frixione-Kunszt-Signer subtraction scheme~\cite{Frixione:1995ms} for 
soft and collinear real radiation with the decomposition into sectors to treat 
real radiation at NNLO, 
as proposed recently in \cite{Czakon:2010td}, is also promising with regards
to the reduction of the number of functions produced by the decomposition.
A combination of sector decomposition with non-linear variable transformations as proposed in 
\cite{Anastasiou:2010pw} can also serve to reduce considerably 
the number of functions to integrate, but is less straightforward to automate completely.

To date, the method of sector decomposition 
has been applied successfully to a considerable number 
of higher order calculations, for a review we refer to  \cite{Smirnov:2006ry,Heinrich:2008si}.
Here we will concentrate on the method of sector decomposition from a 
programming point of view.

Despite its success in practical applications, for quite some time
there was no formal proof for the existence of a
strategy for the iterated sector decomposition  such that 
the iteration is always guaranteed to terminate. This gap has been filled in Ref.~\cite{Bogner:2007cr}, 
by mapping the problem to Hironaka's Polyhedra game~\cite{Hironaka}
and offering three strategies which are proven to terminate.
Bogner and Weinzierl also implemented the algorithm in a public computer program 
for iterated sector decomposition written in C++~\cite{Bogner:2007cr}.
A Mathematica interface to this program, which also allows the calculation of contracted tensor integrals, 
 has recently been published in \cite{Gluza:2010rn}.

A different strategy guaranteed to terminate, leading to less 
subsectors than the strategies of Ref.\,\cite{Bogner:2007cr}, 
was given by A.V.~Smirnov and
M.~Tentyukov, who implemented the algorithm 
in the public program {\small FIESTA}\,\cite{Smirnov:2008py}.  
Based on a detailed analysis of Hepp and Speer sectors in
Ref.\,\cite{Smirnov:2008aw}, an alternative strategy, 
which is based on Speer sectors, has been implemented in 
{\tt FIESTA2}~\cite{Smirnov:2009pb}. 
As the latter strategy also uses 
information on the topology of the graph, it can perform the decomposition more 
efficiently in certain cases.

Another group has implemented~\cite{Ueda:2009xx} the sector decomposition
algorithm in {\tt FORM}\,\cite{Vermaseren:2000nd}. 
Mapping sector decomposition  to convex geometry 
and using algorithms in computational geometry
lead to a guaranteed terminating strategy which seems to be optimal 
with regards to the number of produced
 subsectors~\cite{Kaneko:2009qx,Kaneko:2010kj}.

Compared to the existing packages, the present program for sector decomposition has 
several new features, the main ones being a sophisticated treatment of 
(potential) numerical instabilities and the possibility to apply the program to more 
general functions, like e.g. parameter integrals occurring 
in phase space integrals over real radiation matrix elements, 
or functions including symbolic parameters. 
Half-integer powers of polynomial functions are also allowed.

The structure of the article is as follows. In section \ref{sec:th}, we 
briefly review the theoretical framework.
Then we give an overview of the program structure and explain
the individual components. The installation and usage 
of the program are described in section \ref{sec:install}, followed by
a number of  examples illustrating different aspects of the program in section \ref{sec:demos}. 
The appendix contains more details about timings and phase space parametrisations as well as a 
section with further information for the user.

\section{Theoretical background}
\label{sec:th}
\subsection{Feynman integrals}
A general Feynman graph $G^{\mu_1\ldots\mu_R}_{l_1\ldots l_R}$ in $D$ dimensions 
at $L$ loops with  $N$ propagators  
and $R$ loop momenta in the numerator, where 
the propagators can have arbitrary, not necessarily integer powers $\nu_j$,  
has the following representation in momentum space:
\begin{eqnarray}\label{eq0}
G^{\mu_1\ldots \mu_{R}}_{l_1\ldots l_R} &=& \int\prod\limits_{l=1}^{L} \rd^D\kappa_l\;
\frac{k_{l_1}^{\mu_1}\ldots k_{l_R}^{\mu_R}}
{\prod\limits_{j=1}^{N} P_{j}^{\nu_j}(\{k\},\{p\},m_j^2)}\nn\\
\rd^D\kappa_l&=&\frac{\mu^{4-D}}{i\pi^{\frac{D}{2}}}\,\rd^D k_l\;,\;
P_j(\{k\},\{p\},m_j^2)=q_j^2-m_j^2+i\delta\;,
\end{eqnarray}
where the $q_j$ are linear combinations of external momenta $p_i$ and loop momenta $k_l$.
Introducing Feynman parameters leads to
\bea
G^{\mu_1\ldots \mu_{R}}_{l_1\ldots l_R}&=&  
\frac{\Gamma(N_\nu)}{\prod_{j=1}^{N}\Gamma(\nu_j)}
\int_0^\infty \,\prod\limits_{j=1}^{N}\rd x_j\,\,x_j^{\nu_j-1}\, 
\delta\big(1-\sum_{i=1}^N x_i\big)
\int \rd^D\kappa_1\ldots\rd^D\kappa_L\nn\\
&&k_{l_1}^{\mu_1}\ldots
k_{l_R}^{\mu_R}\,\left[ 
       \sum\limits_{i,j=1}^{L} k_i^{\rm{T}}\, M_{ij}\, k_j  - 
       2\sum\limits_{j=1}^{L} k_j^{\rm{T}}\cdot Q_j +J +\idel
                             \right]^{-N\nu}\;,\label{EQ_mixed_rep}
\eea 
where $N_\nu=\sum_{j=1}^N\nu_j$, $M$ is a $L\times L$ matrix containing Feynman parameters,
$Q$ is an $L$-dimensional vector composed of external momenta and 
Feynman parameters, and $J$ contains kinematic invariants and Feynman parameters.

To perform the integration over
the loop momenta $k_l$, we perform the following shift in order to obtain a quadratic
form for the term in square brackets in eq.~(\ref{EQ_mixed_rep}):
\be
k_l^{\prime}=k_l-v_l\;,
\quad
v_l=\sum_{i=1}^L M^{-1}_{li} Q_i\;.
\ee
After momentum integration one obtains
\begin{eqnarray}\label{EQ:param_rep}
G^{\mu_1\ldots \mu_{R}}_{l_1\ldots l_R} &=& (-1)^{N_{\nu}}
\frac{1}{\prod_{j=1}^{N}\Gamma(\nu_j)}\int
\limits_{0}^{\infty} 
\,\prod\limits_{j=1}^{N}dx_j\,\,x_j^{\nu_j-1}\,\delta(1-\sum_{l=1}^N x_l)\nonumber\\
&&\sum\limits_{m=0}^{\lfloor R/2\rfloor}(-\frac{1}{2})^m\Gamma(N_{\nu}-m-LD/2)
\left[(\tilde M^{-1}\otimes g)^{(m)}\,\tilde l^{(R-2m)}
\right]^{\Gamma_{1},\ldots,\Gamma_{R}}\nonumber\\
&&\times\,\frac{{\cal U}^{N_{\nu}-(L+1) D/2-R}}
{{\cal F}^{N_\nu-L D/2-m}}\\
 &&\nonumber\\
\mbox{where} \qquad \quad &&\nonumber\\
{\cal F}(\vec x) &=& \det (M) 
\left[ \sum\limits_{j,l=1}^{L} Q_j \, M^{-1}_{jl}\, Q_l
-J -\idel\right]\label{DEF:F}\\
{\cal U}(\vec x) &=& \det (M) \nonumber\\
\tilde M^{-1}&=&{\cal U}M^{-1}\;,\quad
\tilde l={\cal U}\,v\nn
\end{eqnarray}   
and $\lfloor R/2\rfloor$ denotes the nearest integer less or equal to $R/2$. 
The expression \\
$[(\tilde M^{-1}\otimes g)^{(m)}\,\tilde l\,^{(R-2m)}]^
{\Gamma_{1},\ldots,\Gamma_{R}}$ stands for the sum over all different combinations 
of $R$ double-indices distributed to $m$ metric tensors and $(R-2m)$ 
vectors $\tilde l$. The double indices 
$\Gamma_{i}=(l,\mu_i(l))\,,\,l\in\{1,\ldots,L\}, i\in \{1,\dots ,R\}$
denote the $i^{\rm{th}}$ Lorentz index, belonging to the $l^{\rm{th}}$ loop momentum. 

\vspace*{3mm}

As can be seen from Eq.~(\ref{EQ:param_rep}), the difference between 
scalar ($R=0$) and tensor ($R>0$) integrals, once  the Lorentz structure is extracted, is
given by the fact that there are additional polynomials of Feynman parameters in 
the numerator. These polynomials can simply be included into 
the sector decomposition procedure, thus treating contracted tensor integrals
directly without reduction to scalar integrals.

The functions ${\cal U}$ and ${\cal F}$ also can be constructed
from the topology of the corresponding Feynman graph.
For more details we refer to~\cite{nakanishi,Tarasov:1996br,Smirnov:2006ry,Heinrich:2008si}.

${\cal U}$ is a positive semi-definite function. 
Its vanishing is related to the  UV subdivergences of the graph. 
Overall UV divergences, if present,
will always be contained in the  prefactor $\Gamma(N_{\nu}-m-LD/2)$. 
In the region where all invariants formed from external momenta are negative, 
which we will call the Euclidean region in the following, 
${\cal F}$ is also a positive semi-definite function 
of the Feynman parameters $x_j$.  Its vanishing does not necessarily lead to 
an IR singularity. Only if some of the invariants are zero, 
for example if some of the external momenta
are light-like, the vanishing of  ${\cal F}$  may induce an IR divergence.
Thus it depends on the {\em kinematics}
and not only on the topology (like in the UV case) 
whether a zero of ${\cal F}$ leads to a divergence or not. 
The necessary (but not sufficient) conditions for an IR divergence 
are given by the Landau equations~\cite{Landau:1959fi,ELOP,Tkachov:1997ap}, 
which, in parameter space, simply mean that 
the necessary condition ${\cal F}=0$ for an IR divergence can only 
be fulfilled if some of the parameters $x_i$ go to zero, provided that 
all kinematic invariants formed by external momenta are negative.

For a diagram with massless propagators, 
none of the Feynman parameters occurs quadratically in 
the function ${\cal F}={\cal F}_0$ . If massive internal lines are present, 
${\cal F}$ gets an additional term 
${\cal F}(\vec x) =  {\cal F}_0(\vec x) + {\cal U}(\vec x) \sum\limits_{j=1}^{N} x_j m_j^2$. 
If the power of the Feynman parameters in the polynomial forming ${\cal F}$ 
is larger than one for at least two different  parameters, 
initially or at a later stage in the iterated decomposition, an infinite recursion can occur.  
This happens in the example given in section \ref{sec:QEDbox}
if the ``naive" decomposition strategy is employed.
We have implemented a heuristic procedure to change to a different decomposition strategy 
only in cases where at least two Feynman parameters occur quadratically, which lead to 
a terminating algorithm without producing a large number of subsectors.
We do not claim that this procedure is guaranteed to terminate, 
but it proved useful for practical purposes.

\subsection{General parameter integrals}

The program can also deal with parameter integrals which are  more 
general than the ones related to multi-loop integrals. 
The only restrictions are that the integration domain should be the unit hypercube, 
and the singularities should be only endpoint singularities, i.e. should be located at zero 
or one.
The general form of the integrals is 
\be
I=\int_0^1 dx_1 \ldots \int_0^1 dx_N \prod_{i=1}^m P_i(\vec{x},\{\alpha\})^{\nu_i}\;,
\label{eq:general}
\ee
where $P_i(\vec{x},\{\alpha\})$ are polynomial functions of the parameters $x_j$,
which can also contain some symbolic constants $\{\alpha\}$. 
The user can leave the parameters $\{\alpha\}$ symbolic during the decomposition,
specifying numerical values only for the numerical integration step.
This way the decomposition and subtraction steps do not have to be redone if 
the values for the constants are changed.
The $\nu_i$ are powers of the form $\nu_i=a_i+b_i\epsilon$ 
(with  $a_i$ such that the integral is convergent; note that half integer powers are also possible).
We would like to point out that most phase space integrals in $D$ dimensions over real radiation 
matrix elements can also be remapped to functions of the type (\ref{eq:general}). 
Examples are given in section \ref{sec:demos}.

\subsection{Iterated sector decomposition}\label{itersd}

Here we will review the basic algorithm of iterated sector decomposition 
only briefly. For more details we refer to \cite{Binoth:2000ps,Heinrich:2008si}.

Our starting point is a function of the form of Eq.~(\ref{eq:general}).
Loop integrals in the form of Eq.~(\ref{EQ:param_rep}),
with open Lorentz indices contracted by external momenta or metric tensors,
are a special case thereof,
distinguished by the presence of a $\delta$-distribution which constrains the sum of the
integration parameters.  Therefore loop integrals are treated somewhat differently
than the more general functions, 
meaning that the $\delta$-constraint is integrated out in a special way leading to 
the so-called {\it primary sectors}. 

 \subsubsection*{I.  Generation of primary sectors (loop integrals only)}
 \label{sec:primary}

We split the integration domain into
$N$ parts and eliminate the $\delta$-distribution in such a way that the remaining 
integrations are from 0 to 1.  
To this aim we decompose the integration range into $N$ 
sectors, where in each sector $l$, $x_l$ is largest:
\begin{eqnarray}
\int_0^{\infty}d^N x =
\sum\limits_{l=1}^{N} \int_0^{\infty}d^N x
\prod\limits_{\stackrel{j=1}{j\ne l}}^{N}\theta(x_l - x_j)\;.
\end{eqnarray} 
The integral is now split into $N$ domains corresponding   
to $N$ integrals $G_l$ from which we extract a common factor:
 $G=(-1)^{N_\nu} \Gamma(N_\nu-LD/2) \sum_{l=1}^{N} G_l$. 
 The $N$ generated sectors
will be called {\em primary} sectors in the following.
 In the  integrals $G_l$ we substitute 
\begin{eqnarray}
x_j = \left\{ \begin{array}{lll} x_l x_j     & \mbox{for} & j<l \\
                                   x_l         & \mbox{for} & j=l \\
                                   x_l x_{j-1} & \mbox{for} & j>l \end{array}\right.
\end{eqnarray} 
and then  integrate out $x_l$ using the $\delta$-constraint.
As ${\cal U},{\cal F}$ are homogeneous of degree $L$,\,$L+1$, respectively,  
and  $x_l$ factorises completely, we have
${\cal U}(\vec x) \rightarrow {\cal U}_l(\vec t\,)\, x_l^L$ and
${\cal F}(\vec x) \rightarrow {\cal F}_l(\vec t\,)\, x_l^{L+1}$
and thus, using $\int dx_l/x_l\,\delta(1-x_l(1+\sum_{k=1}^{N-1}x_k ))=1$, we obtain 
\begin{eqnarray}\label{EQ:primary_sectors}
 G_l &=& \int\limits_{0}^{1} \prod_{j=1}^{N-1}{\rd}x_j\,x_j^{\nu_j-1}
\frac{ {\cal U}_l^{N_\nu-(L+1)D/2}(\vec{x}\,)}{ {\cal F}_l^{N_\nu-L D/2}(\vec{x}\,)} 
\quad , \quad l=1,\dots, N \;.
\end{eqnarray} 
Note that the singular behaviour leading to $1/\epsilon$--poles
still comes  from regions where a set of parameters $\{x_j\}$ 
goes to zero. 

\subsubsection*{II.  Extraction of the singular factors}
\label{sec:iter}

For functions of the form Eq.~(\ref{eq:general}), we first have to determine 
which of the Feynman parameters generate singularities at $x_j=1$, and which ones 
can lead to singularities at zero {\it and} one.
The parameters $x_j$ for which a denominator vanishes at  $x_j=1$ but not at $x_j=0$
should be  remapped by the transformation $x_j\to 1-x_j$. 
If the integrand can become singular at both endpoints 
of the integration range for a parameter $x_j$,   we  split the integration range at 1/2:
After the split 
\be
\int_0^1 {\rm d} x_j=\underbrace{\int_0^\frac{1}{2} {\rm d} x_j}_{(a)}+
\underbrace{\int_\frac{1}{2}^1 {\rm d} x_j}_{(b)}
\label{eq:split}
\ee
and the substitution $x_j=z_j/2$ in $(a)$ and $x_j=1-z_j/2$ in $(b)$, 
all endpoint singularities occur at $z_j\to 0$ only.
This splitting is done automatically  by the program; the user only has to define which variables 
should be split.

\medskip

At this stage our starting point is a parametric integral where the integrand vanishes
if some of the integration parameters go to zero. 
Our aim is to factorise the singularities, i.e. extract them in terms of overall factors
of type $x_j^{a_j+b_j\epsilon}, \,a_j\leq -1$.
We proceed as follows.

\begin{description}
\item[1.] Determine a minimal set of parameters, say 
${\cal S}=\{x_{\beta_1},\dots ,x_{\beta_r}\}$, such that  
at least one of the functions $P_i(\vec{x},\{\alpha\})$ vanishes 
if the parameters of ${\cal S}$ are set to zero. 
\item[2.] The corresponding integration range is an $r$-cube 
which is decomposed into $r$ {\em subsectors}
by decomposing unity according to
\begin{eqnarray}
\prod\limits_{j=1}^r \theta(1 - x_{\beta_j}\geq  0)\,\theta(x_{\beta_j})=
\sum\limits_{k=1}^r \prod\limits_{\stackrel{j=1}{j\ne k}}^r 
\theta(x_{\beta_k}- x_{\beta_j}\geq 0)\,\theta(x_{\beta_j})\;.
\end{eqnarray}
 \item[3.] Remap  the variables to the unit hypercube in each new 
 subsector by the substitution
 \begin{eqnarray}
x_{\beta_j} \rightarrow 
\left\{ \begin{array}{lll} x_{\beta_k} x_{\beta_j} &\mbox{for}&j\not =k \\
                           x_{\beta_k}              &\mbox{for}& j=k\,.  \end{array}\right.
\end{eqnarray}
This gives a Jacobian factor of $x_{\beta_k}^{r-1}$. By construction
$x_{\beta_k}$ factorises from at least one of the functions $P_i(\vec{x},\{\alpha\})$.
\end{description}
For each subsector the above steps have to be repeated 
as long as a set ${\cal S}$ 
can be found such that one of the rescaled functions $\tilde{P}_i(\vec{x},\{\alpha\})$
vanishes 
if the elements of ${\cal S}$ are set to zero. 
This way  new subsectors are created in each subsector 
of the previous iteration, resulting in a 
tree-like structure after a certain number of iterations. 
The iteration stops if the functions 
$\tilde{P}_i(\vec{x},\{\alpha\})$ 
contain a constant term, i.e. if they are of the  form 
\begin{eqnarray}\label{EQ:subsec_UF}
\tilde{P}_i(\vec{x},\{\alpha\}) &=& \alpha_0 +  \tilde{Q}_i(\vec x\,,\{\alpha\}) \;,
\end{eqnarray}
where
$\tilde{Q}_i(\vec x,\{\alpha\})$  are polynomials in the
variables $x_j$, and $\alpha_0$ is a constant, i.e. $\lim_{\vec{x}\to 0}\tilde{Q}_i(\vec x,\{\alpha\})$ is nonzero.

\medskip

The resulting subsector integrals have the general form
\begin{eqnarray}\label{EQ:subsec_form}
I &=& \int\limits_{0}^{1} \left( \prod_{j=1}^{N}{\rd}x_j
\; x_j^{a_j+b_j\epsilon}  \right)
\prod_{i=1}^m \tilde{P}_i(\vec{x},\{\alpha\})^{\nu_i}\;.
\end{eqnarray}
The singular behaviour of the integrand now can be 
read off directly from the exponents $a_j$, $b_j$ for a given subsector integral. 

\subsubsection*{III.  Subtraction of the poles}

For a particular $x_j$, the integrand after the factorisation described above, is of the form
\begin{eqnarray}\label{EQsub_step1}
I_j = \int\limits_0^1 dx_j\, x_j^{a_j+b_j\epsilon}\, {\cal I}(x_j,\{x_{i\not=j}\},\epsilon) \;.
\end{eqnarray}
If  $a_j > -1$,  no subtraction is needed and one can go to the next
variable $x_{j+1}$. If  $a_j \leq -1$, one expands ${\cal I}(x_j,\{x_{i\not=j}\},\epsilon)$ into a 
Taylor series around $x_j=0$. 
Subtracting the Taylor series (to order\footnote{To account for half-integer exponents, e.g. 
$a_j=-3/2$, we use $\lfloor|a_j|\rfloor$, denoting the nearest integer less or equal to $|a_j|$.} 
$p$ for $|a_j|=p+1$) and adding it back in integrated form, 
we obtain a part where the poles are subtracted and a part exhibiting $1/\eps$ poles times a function 
depending only on the remaining integration parameters. 
\begin{eqnarray}\label{EQsub_step2}
&&I_j = \sum\limits_{p=0}^{\lfloor |a_j|\rfloor-1} \frac{1}{a_j+p+1+b_j\epsilon}
 \frac{{\cal I}_j^{(p)}(0,\{x_{i\not=j}\},\epsilon)}{p!} 
+ \int\limits_{0}^{1} dx_j \, x_j^{a_j+b_j \epsilon}\, R(\vec{x},\epsilon) \nn\\
&&R(\vec{x},\epsilon)={\cal I}(\vec{x},\eps)-\sum\limits_{p=0}^{\lfloor |a_j| \rfloor-1}
{\cal I}_j^{(p)}(0,\{x_{i\not=j}\},\epsilon)\,\frac{x_j^p}{p!}
\;.
\label{tjsubtr}
\end{eqnarray}
For $a_j=-1$, expanding the above expression in $\epsilon$ is
equivalent to an expansion in ``plus distributions"\,\cite{Gelfand,Anastasiou:2003gr}
\begin{eqnarray}
\label{plusdist}
&&x^{-1+b\,\epsilon}=\frac{1}{b\,
\epsilon}\,\delta(x)+
\sum_{n=0}^{\infty}\frac{(b\,\epsilon)^n}{n!}
\,\left[\frac{\ln^n(x)}{x}\right]_+\;,\nn\\
&&\mbox{where }\nn\\
&&\int_0^1 dx \,f(x)\, \left[g(x)/x\right]_+=\int_0^1 dx \, 
\frac{f(x)-f(0)}{x}\,g(x)\;,
\end{eqnarray}
with the integrations over the terms containing $\delta(x)$ already
carried out.

After having done the subtractions for each $x_j$, 
all poles are extracted, such that the resulting 
expression can be expanded in $\epsilon$. This defines
a Laurent series in $\epsilon$
\begin{eqnarray}\label{EQ:eps_series_Glk}
 I = \sum\limits_{n=-LP}^{r} C_{n}\,\epsilon^n + 
{\cal O}(\epsilon^{r+1}) \;,
\end{eqnarray}
 where the coefficients are finite parameter integrals
 of  dimension $(N-1-|n|)$ for $n< 0$ and of dimension $(N-1)$ for $n\geq 0$.
 $LP$ denotes the leading pole, which can be at most $2L$ for an $L$-loop integral.
 The finite coefficient functions can be integrated by Monte Carlo integration 
 if the Mandelstam invariants in ${\cal F}$ respectively the 
 numerical constants in a general integrand have been chosen such that 
 the integrand does not vanish in the integration domain.
 
 \subsubsection*{Improving the numerical stability}
 
For $a_j=-1$ in eq.~(\ref{EQsub_step2}), the singularity is of logarithmic nature, 
i.e. $\sim \log(\Lambda)$ if a lower cutoff $\Lambda$ for the parameter integral was used.
In renormalisable gauge theories, linear ($a_j=-2$) 
or even higher ($a_j<-2$) poles should not occur.
However, they can occur at intermediate stages
of a calculation, and as they are formally regulated by dimensional regularisation, 
a method has been worked out for the program to be able to deal with  higher than logarithmic 
singularities efficiently.
This method relies on integration by parts (IBP) in a way which aims at maximal numerical stability.

Let us consider an integrand {\it after} subtraction, as the one in eq.~(\ref{EQsub_step2}), 
and focus on only one variable, $x_j=x$, and define 
\be
I(R,a,b)=\int\limits_{0}^{1} dx \, x^{a+b \epsilon}\, R( x,\epsilon) \;,
\ee
where, in the case $a<0$,  $R(x,\eps)$ is ${\cal O}(x^{-a})$ as $x\to{0}$ by construction.
Integration by parts gives
\bea
I_{\rm{BP}}(R,a,b)&=&\frac{1}{a+1+b\epsilon}\,\left\{\left[x^{a+1+b\epsilon}R(x,\eps)\right]^1_0
-\int^1_0 dx\,x^{a+1+b\epsilon}R'(x)\right\}\label{ibp1}\\
&=&\frac{1}{a+1+b\epsilon}\left\{R(1,\eps)
-I(R',a+1,b)\right\}\;.\label{ibp2}
 \eea
 As $x^{a+1+b\epsilon}R(x,\eps)$ vanishes as $x\to{0}$, the term $\sim R(0,\eps)$ in the square bracket in 
 eq.~(\ref{ibp1}) is zero.\\
 Also notice that $R'(x)\sim x^{-a-1}$ as $x\to{0}$, so $I(R',a+1,b)$ can be treated in the same way.
 
 For the case $a=-1$ we thus obtain, expanding in $\epsilon$
 \bea
 I_{\rm{BP}}(R,-1,b)&=&-\frac{1}{b\epsilon}\,\int^1_0 dx\,R'(x)\,
 \sum _{n=1}^{\infty}\frac{(b\,\epsilon)^n}{n!}
\,\ln^n(x)\;.
 \eea
 For $a<-1$, we can use eq.~(\ref{ibp2}) to iterate this procedure until we
 reach $a>=-1$.
 
 This method is certainly beneficial in the case of numerical instabilities coming from terms of the form 
 $[f(x)-f(0)]/x$ or $[f(x)-f(0)-x\,f'(0)]/x^2$ in $R(x,\eps)$, leading to differences of large numbers for 
 $x\to 0$.
 The IBP method raises the powers of $x$, trading them for 
 additional logarithms in the integrand, which can be integrated more easily by the Monte Carlo program.
 Note also that the whole procedure is linear in $R(x,\eps)$, so it allows one to split $R(x,\eps)$ 
 into smaller functions which can be dealt with more easily by the numerical integrator.

 \subsubsection*{Error treatment}
 
  We usually integrate the sum of a small set of  functions stemming 
  from the $\epsilon$-expansion of a certain pole structure 
  individually, and afterwards sum all the individual results contributing to a certain pole coefficient.
  The set of functions to sum before integration is defined by the size of the individual functions: 
  the functions are summed until their sum reaches about two Megabytes.
  
  The errors are calculated by adding the Monte Carlo errors of the individual integrations in quadrature.
  It is possible that there are large cancellations between different functions contributing to the same 
  pole coefficient. In such cases it is better to first sum all coefficient functions and then integrate.
  The program offers the possibility to do so as an option which can be specified in the 
  input parameter file.

The user should be aware that for complicated functions 
 containing many subtractions, the Monte Carlo error estimate is not quite appropriate:
 it is calculated on a purely statistical basis, scaling like $1/\sqrt{N}$ if $N$ 
 is the number of sampling points. However, this is only a reliable error estimate 
 under the assumption that the sampling  
 has mapped  all the important features of the function
 (i.e. all peaks)  sufficiently precisely, and strictly is only valid for square integrable functions.
 If the function is not square integrable (but integrable), 
 the Monte Carlo estimate for the integral will still
 converge to the true value, but the error estimate will become unreliable.
 For more involved integrals, we are faced with functions which have gone through
 numerous decompositions and subtractions, such that their shape in the unit hypercube is 
 quite complicated, and therefore the naive Monte Carlo error estimate 
 tends to underestimate the ``true" error. 
 
Often the main source of underestimated errors in the final result is the fact      
that there are a large number of integrations to sum, and so adding the errors      
in quadrature would only give a truly appropriate error estimate if there were      
no systematic errors in the numerical integration. 
 
 We should also note that converting the result to extract a different $\eps$-dependent  prefactor 
 may lead to cancellations between different contributions to a certain pole coefficient
 such that the error estimate may be too optimistic in these cases.

\section{Structure of the program}

\begin{figure}[htb]
\includegraphics[width=15.cm]{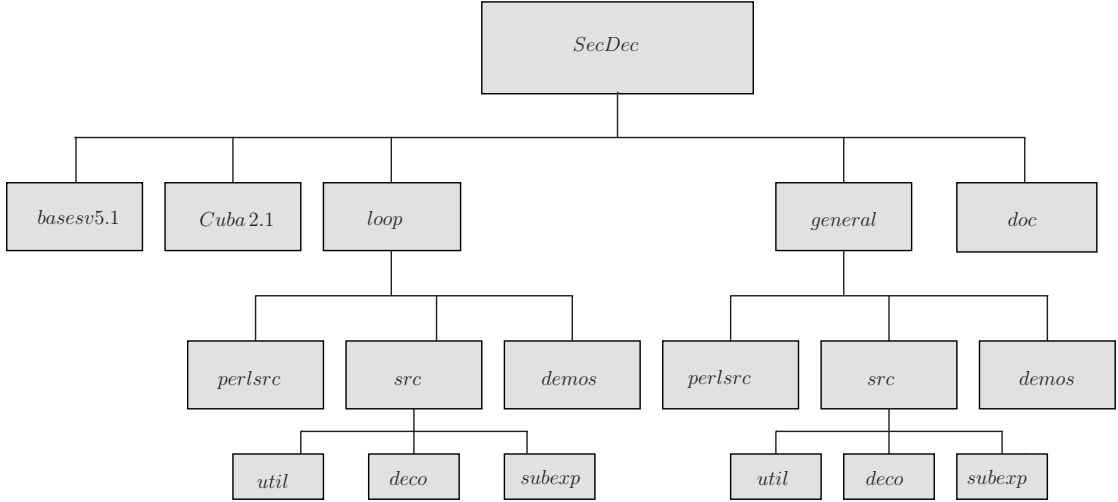}
\caption{Directory structure of the SecDec program.}
\label{fig:structure}
\end{figure}

\begin{figure}[htb]
\includegraphics[width=15cm]{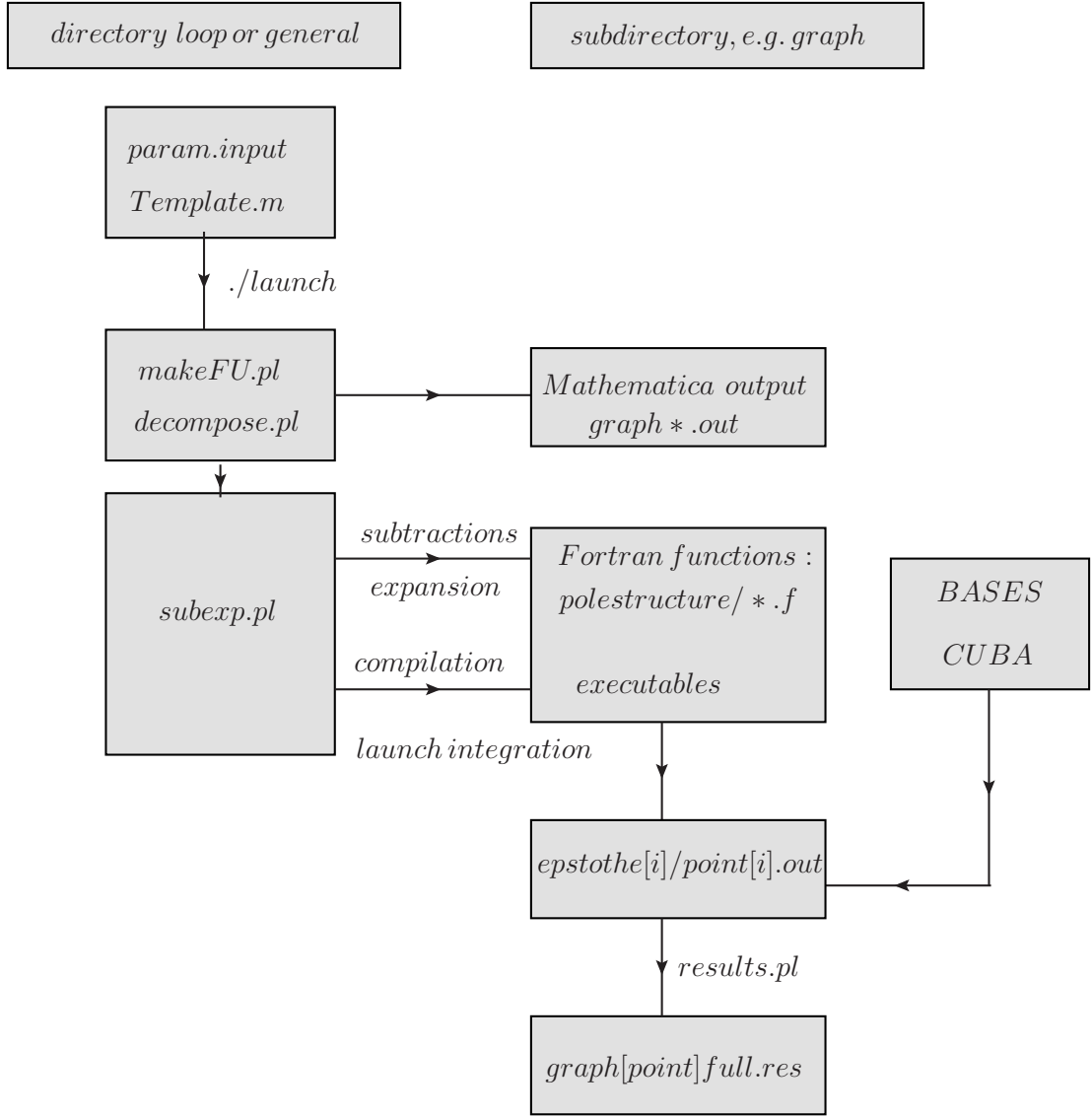}
\caption{Flowchart showing the main steps the program 
performs  to produce the result files. In each of the subdirectories 
{\tt loop} or {\tt general}, the file {\tt Template.m} can be used to define the integrand. 
The produced files are written to a subdirectory created according to the settings given 
in {\tt param.input}. By default, a subdirectory with the name of the graph or integrand is created to store the produced functions. 
This directory will contain subdirectories according to the pole structure of the integrand.
The perl scripts (extension .pl) are steering the various steps to be performed by the program.}
\label{fig:flowchart}
\end{figure}

The program consists of two parts, an algebraic part and a numerical part.
The algebraic part uses code written in Mathematica~\cite{Wolfram} and does 
the decomposition into sectors, the subtraction of the 
singularities, the expansion in $\eps$ and the generation of the 
files necessary for the numerical integration. 
In the numerical part,  Fortran functions forming the coefficient 
of each term in the Laurent series in $\eps$ are integrated using the 
Monte Carlo integration program {\small BASES}, version 5.1\,\cite{Kawabata:1995th}, 
or one of the routines from the {\small CUBA} library, version 2.1\,\cite{Hahn:2004fe}.
The different subtasks are handled by perl scripts. 
The directory structure of the program is shown in Fig.~\ref{fig:structure}, 
while the  flowchart in Fig.~\ref{fig:flowchart} shows the basic 
flow of input/output streams.

The  directories {\tt loop} and {\tt general} have the same global structure, 
only some of the individual files are specific to loops or to more general parametric functions.
The directories contain a number of perl scripts steering the 
decomposition and the numerical integration.  The scripts use perl modules contained 
in the subdirectory {\tt perlsrc}.

The Mathematica source files 
are located in the subdirectories {\tt src/deco}: files used for the decomposition, 
{\tt src/subexp}: files used for the pole subtraction and expansion in $\eps$, 
{\tt src/util}: miscellaneous useful functions.  
For the translation of the Mathematica expressions to Fortran77 functions we 
use the package {\tt Format.m}\,\cite{Format}.
The subdirectories {\tt basesv5.1} and {\tt Cuba-2.1} contain 
the libraries for the numerical integration, 
taken from \cite{Kawabata:1995th} and \cite{Hahn:2004fe}, respectively.
The documentation, created by {\it robodoc}\,\cite{robodoc} 
is contained in the subdirectory {\tt doc}. It contains an index to look up documentation 
of the source code in html format by loading {\tt masterindex.html} into a browser.

\begin{figure}
\includegraphics[width=14.2cm]{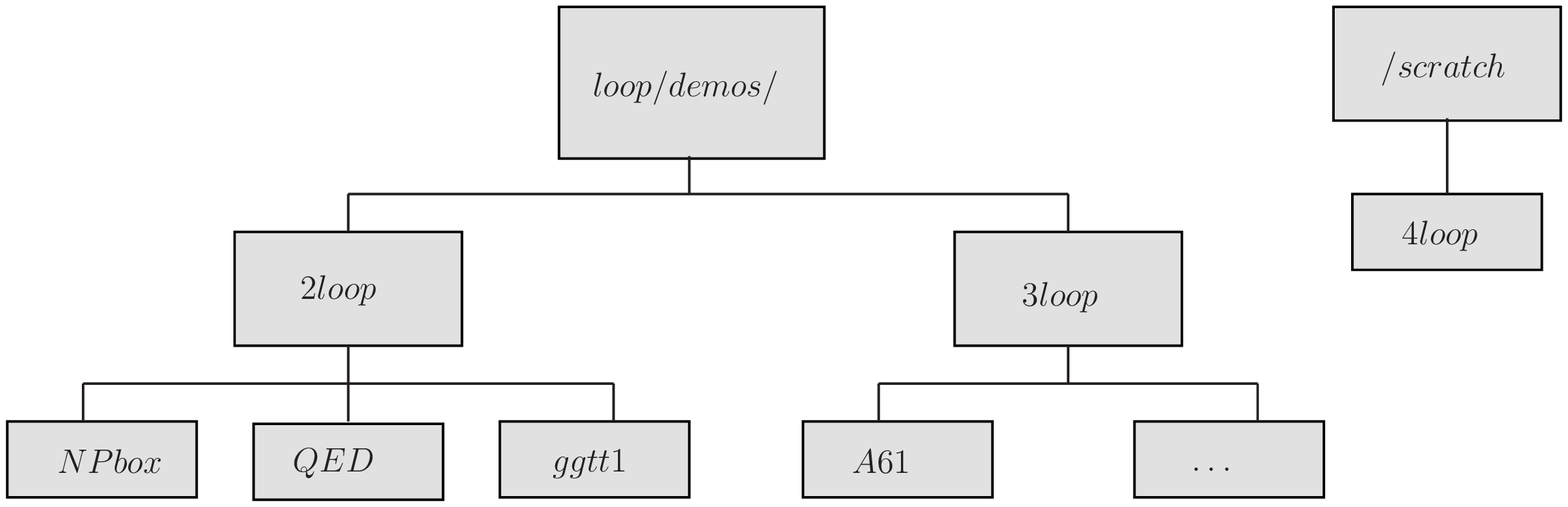}
\caption{Example for a directory structure created by running the loop demo programs 
NPbox, QED, ggtt1, A61. A  four-loop example defined by the user to be written to the scratch 
disk is also shown.
\label{fig:UserDirstructure}}
\end{figure}

The intermediate files and the results will be stored in a subdirectory of the working directory
whose name {\it mysubdir} can be specified by the user (first entry in {\tt param.input}, 
leaving this blank is a valid option).
A subdirectory of {\it mysubdir} with the name of the graph, respectively integral to calculate 
will be created by default.
If the user would like to store the files in a directory which is not the subdirectory of the working directory, for example in {\it /scratch}, he can do this by specifying the full path in the second entry in {\tt param.input}. 
An example of a  directory structure created by running the examples {\it NPbox, QED, ggtt1, A61},
a user-defined 3-loop example, and a 4-loop example to be written to the scratch disk is given in Fig.~\ref{fig:UserDirstructure}.

\begin{figure}
\includegraphics[width=14.cm]{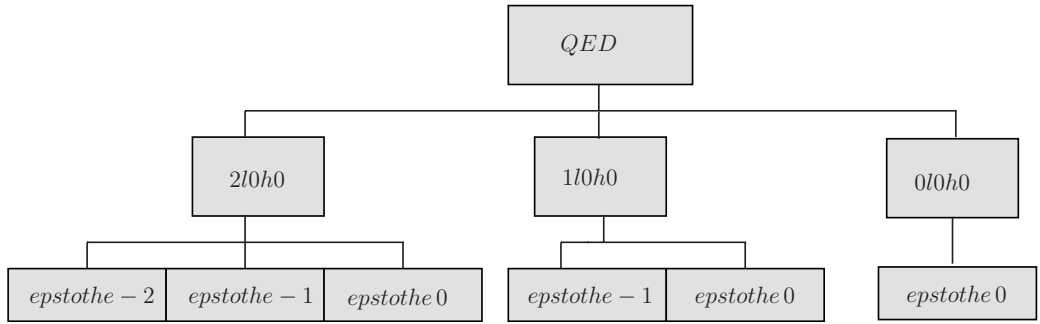}
\caption{Example for a directory tree corresponding to the pole structure of 
the graph {\tt QED} contained in the demo programs.
\label{fig:polestructure}}
\end{figure}

The directory created for each graph will contain subdiretories according to the pole structure 
of the graph. The labelling for the  pole structure is of the form e.g. {\tt 2l0h0}, 
denoting 2 logarithmic poles, no linear and no higher poles. 
We should point out that this labelling does not necessarily correspond to the 
final pole structure of the integral. It is merely for book-keeping purposes, 
and is based on the counting of the powers of the factorised integration variables.
In more detail, if $i_1$ variables have power $-1$, $i_2$ variables have a power 
$-2\leq i_2 < -1$ and $i_3$ variables have a power $<-2$, the labelling will be 
{\tt $i_1\,l\,i_2\,h\,i_3$}, even though the non-logarithmic poles will disappear 
upon $\eps$-expansion. In particular, for half-integer powers, the labelling does not 
correspond to  ``true" poles,  but rather to terms which can be cast into functions like 
$\Gamma(-3/2-\eps)$, which are well-defined in the context of dimensional regularisation, 
where $\eps$ can be regarded as an arbitrary 
(complex) parameter.
Note also that in the case of a prefactor containing $1/\eps$ poles multiplying the parameter 
integral, the poles which are flagged up at this stage of the program will only correspond to 
the poles read off from the integration parameters.
In any case, the final result will be given to the order specified by the user 
in {\tt param.input}. \\
Each of these ``polestructure" directories contains 
further subdirectories where the files for a particular power in epsilon are stored.
An example is given in Fig.~\ref{fig:polestructure}.

The user only has to edit the following two files:
\begin{itemize}
\item {\tt param.input}: (text file)\\
specification of paths, type of integrand, 
order in $\eps$, output format, parameters for numerical integration, 
further options
\item {\tt Template.m}: (Mathematica syntax)
\begin{itemize}
\item for loop integrals: specification of loop momenta, propagators; optionally numerator, non-standard propagator powers
\item for general functions: specification of integration variables, integrand, variables to be split
\end{itemize}
\end{itemize}
To give a specific example rather than empty templates, the files {\tt param.input} and 
{\tt Template.m} in the {\tt loop} subdirectory contain the setup for example 1, 
the non-planar massless on-shell two-loop box diagram, while those in the {\tt general} 
directory contain the setup for example 6, a hypergeometric function of type $\,_5F_4$.

Apart from these default parameter/template files, the program comes with example 
input and template files in the subdiretories {\tt loop/demos} 
respectively {\tt general/demos}, described in detail in section \ref{sec:demos}.

The user can choose the numerical integration routine and the settings for the different integrators
contained in the Cuba library  in the file {\tt param.input}.  
The compilation of the chosen integration routine with the corresponding settings
will be done automatically by the program.

\section{Installation and usage}\label{sec:install}

\subsection*{Installation}
The program can be downloaded from 
{\tt http://projects.hepforge.org/secdec/}.

Installation is done by unpacking the tar archive, using the command
{\it  tar xzvf SecDec.tar.gz}. 
This will create a directory called {\tt SecDec} 
with the subdirectories as described above. Change to the {\tt SecDec} directory
and run {\it ./install}.

Prerequisites are Mathematica, version 6 or above, perl (installed by default on 
most Unix/Linux systems) and a Fortran compiler (e.g. gfortran, ifort).
The install script only checks if Mathematica and perl are installed on the system and 
inserts the corresponding path into the perl scripts. The install script 
does not test the existence of a Fortran compiler because the compiler should 
be specified by the user in {\tt param.input}. If no compiler is specified, it defaults to 
{\em gfortran}.

\subsection*{Usage}

\begin{enumerate}

\item Change to the subdirectory {\tt loop} or {\tt general}, depending 
on whether you would like to calculate a loop integral or a more general parameter integral.
\item Copy the files {\tt param.input} and {\tt Template.m} to 
create your own parameter and template files  {\tt myparamfile}, {\tt mytemplatefile}.
\item Set the desired parameters in {\tt myparamfile} and define 
the propagators etc. in {\tt mytemplatefile}.
   
\item Execute the command {\it ./launch -p myparamfile -t mytemplatefile} 
in the shell.  \\
If you omit the option {\it -p myparamfile}, the file {\tt param.input} will be taken as default.
Likewise, if you omit the option {\it -t mytemplatefile}, 
the file {\tt Template.m} will be taken as default.
If your files {\it myparamfile, mytemplatefile} are in a different directory, say, 
{\it myworkingdir}, 
 use the option \\{\bf -d myworkingdir}, i.e. the full command then looks like 
 {\it ./launch -d myworkingdir -p myparamfile -t mytemplatefile}, 
 executed from the directory {\tt SecDec/loop} or
 {\tt SecDec/general}. \\
 Alternatively,  you can call the launch script from any directory if you prepend the path
 to the launch script, i.e. 
 the command  \\
 {\it path\_to\_launch/launch -p myparamfile -t mytemplatefile}
 executed from {\bf myworkingdir} would run the program in the same way. 
 {\it path\_to\_launch} can be
 either the full or relative path for {\tt SecDec/loop} or {\tt SecDec/general}.
 
 \vspace{3mm}
 
 The {\it ./launch} command will launch the following  perl scripts:
     \begin{itemize}
    \item {\tt makeFU.pl}: (only for loop integrals)
      constructs the integrand functions ${\cal F}, {\cal U}$ and the numerator function
      from the propagators and indices given in {\tt Template.m}.
    \item {\tt decompose.pl}: 
      launches the iterated sector decomposition
   \item {\tt  subexp.pl}: 
      launches the subtractions and epsilon-expansions and writes the Fortran functions.
      Depending on the ``exe-flag" specified in the parameter file (see below for a detailed explanation
      of the flag), this script also launches the compilation and the numerical integrations.
   \end{itemize} 
   
 \vspace{3mm}

\item Collect the results. Depending on whether you have used a single machine or 
submitted the jobs to a cluster, the following actions will be performed:
 \begin{itemize}
\item If the calculations are done sequentially on a single machine, 
    the results will be collected automatically (via {\tt results.pl} called by {\tt launch}).
    The output file will be displayed with your specified text editor.
    The results are also saved to the files {\tt [graph]\_[point]epstothe*.res}
    and  {\tt [graph]\_[point]full.res} in the subdirectory {\tt subdir/graph} (loops)
    respectively 
    {\tt subdir/integrand} (general integrands)
    (name specified in {\tt param.input}, where you can also specify different names for different numerical points).

\item If the jobs have been submitted to a cluster: 
	when all jobs have finished,  execute  the command 
	{\it ./results.pl [-d myworkingdir -p myparamfile]} 
	in a shell from the directory {\tt SecDec/loop} or 
	{\tt SecDec/general} 
	to create the file containing the final results.

       If the the user needs to change the batch system settings:
        manually edit {\tt perlsrc/makejob.pm} and {\tt perlsrc/launchjob.pm}.
	This writes the desired syntax to the scripts 
	{\tt job[polestructure]} in the corresponding {\tt subdir/graph } or 
	{\tt subdir/integrand} 
	subdirectory. 
 \end{itemize}

 \vspace{3mm}

\item After the calculation and the collection of the results is completed, 
you can use the shell command {\it ./launchclean[graph]}
to remove obsolete files.\\
  If called with no arguments, the script only removes object files, launch scripts, 
   makefiles and executables,  but leaves the Fortran files created by Mathematica, 
  so that different numerical points can be calculated without rerunning the Mathematica code.
If  called with the argument 'all' (i.e. \,{\it ./launchclean[graph]  all}),  
it removes everything except the result files displaying the final result and the timings.

\end{enumerate}

The {\bf 'exe' flag} contained in {\tt param.input} offers the possibility 
 to run the program only up to certain intermediate stages. The flag can take  values  
 from 0 to 4. 
 The different levels are:
 \begin{description}
 \item[exe=0:] does the iterated sector decomposition and writes files 
 containing lists of subsector functions ({\tt graphsec*.out})
                          for each pole structure to the output subdirectory. 
    Also writes  the Mathematica files {\tt subandexpand*.m} for each pole structure, 
                  which serve to do the symbolic subtraction, epsilon expansion 
		  and creation of the Fortran files.
	Also writes the scripts \\{\tt batch[polestructure]} 
		  which serve to launch these jobs at a later stage.
		
 \item[exe=1:] launches the scripts {\tt batch[polestructure]}. 
 This will produce the Fortran functions and write them to individual subdirectories for each pole structure.
 \item[exe=2:] creates all the additional files needed for the numerical integration.
 \item[exe=3:]  compilation is launched to make the executables. 
 \item[exe=4:] the executables are run.
  \end{description}
If the first steps of the calculation, e.g. the decomposition or the creation of the Fortran functions, 
are already done,  the following commands are available to continue the calculation without having to 
restart from scratch:
  \begin{itemize}
\item {\bf finishnumerics.pl [-d myworkingdir -p myparameterfile]: }\\
   if the 'exe' flag in {\tt param.input} resp. {\tt myworkingdir/myparameterfile} is set smaller than four, this 
   will complete the calculation without redoing previous steps. 
  
\item {\bf justnumerics.pl [-d myworkingdir -p myparameterfile]: }\\
if you would like to redo just the numerical integration, for example 
to produce results for a different numerical point or to try out a different number 
of sampling points, iterations etc. for the Monte Carlo integration:
 change  the values for the numerical point resp. the settings for the Monte Carlo  integration
and the ``name of the numerical point"   in the parameter file, 
 and then use the command  
 {\it ./justnumerics.pl [-d myworkingdir -p myparameterfile]} to 
 redo only the numerical integrations  (if the Fortran files f*.f have been produced already).
 Using this option skips the Mathematica subtraction and epsilon expansion step which can be done once and for all,
 as the variables at this stage are still symbolic.
 After completion of the numerical integrations, use the command {\it ./results.pl [-d myworkingdir -p myparameterfile]} to collect and display the results as above.
\end{itemize}

The program tries to detect the path to Mathematica automatically.
In case you get the message ``path for Mathematica not automatically found", 
please insert the path to Mathematica on your system manually for the variable {\tt \$mathpath}
in the file {\tt perlsrc/mathlaunch.pl}.

We also should mention that the code starts working first on the most complicated 
pole structure, which takes longest. 
This is because in case the jobs are sent to a cluster, it is advantageous to 
first send the jobs which are expected to take the most time. 

\section{Description of Examples}\label{sec:demos}

\subsection{Loop integrals}

The examples described below can be found in the subdirectory {\tt loop/demos}.

\subsubsection{Example 1: Non-planar massless two-loop box}
\label{example:npbox}

The non-planar massless two-loop box is a non-trivial example, as the 
sector decomposition applied to the standard representation, produced by 
combining all propagators  simultaneously with Feynman parameters,
exhibits ``non-logarithmic poles" (i.e. exponents of Feynman parameters $\leq -1$)
in the course of the decomposition.
We should point out that, even though  the program can deal with linear or higher poles 
 in a completely automated way, it is often a good idea to 
investigate if the integrand can be re-parametrised  
such that poles of this type do not occur, because these poles 
require complicated subtraction terms which slow down the calculation.
Non-linear transformations as e.g. described in \cite{Anastasiou:2010pw} 
can be useful in this context.
Further,  integrating out first one loop momentum, 
and then combine  the remaining propagators with the obtained intermediate result
using another set of Feynman parameters often 
leads to a representation where at least one of the parameters can be factorised
without sector decomposition, thus speeding up the calculation considerably.
This is demonstrated for the non-planar massless two-loop box in appendix \ref{appendix:xbox}, 
and the template to calculate the graph in this way can be found in {\tt SecDec/general/demos}.

\begin{figure}
\begin{center}
\includegraphics[width=5.5cm]{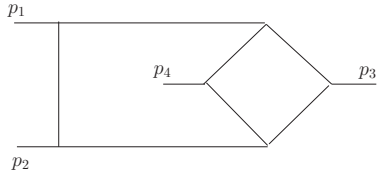}
\end{center}
\caption{The non-planar two-loop box, called {\it NPbox} in example 1.}
\label{fig:np7}
\end{figure}

To obtain results for the non-planar massless two-loop box shown in Fig.~\ref{fig:np7}
without doing any analytical steps, 
copy the file {\tt loop/param.input}  to a new parameter file, say {\tt paramNPbox.input}, 
and  specify the desired order in $\eps$, 
the numerical point and possible further options. Likewise, copy the file {\tt loop/Template.m}  
to a new template file, say {\tt templateNPbox.m}
 (this already has been done for the examples described here, see  the subdirectory 
{\tt loop/demos}).
Then, in the {\tt loop/demos} directory, use the command {\em ../launch -p paramNPbox.input 
-t  templateNPbox.m}.
The file {\tt templateNPbox.m} already has the propagators of the non-planar 
double box predefined.
In {\tt paramNPbox.input}, we defined the prefactor such that a factor of $-\Gamma(3+2\eps)$ 
is {\it not } included in the numerical result: if we define
\be
G_{NP}(s,t,u)=-\Gamma(3+2\eps)\sum_{n=0}^4\frac{P_n}{\eps^n}+{\cal O}(\eps)\;,
\ee
the program should yield the results for $P_n$, given in Table\,\ref{table:np7}. 
Note that according to eq.\,(\ref{eq0}), we always divide $L$-loop integrals by $(i\pi^\frac{D}{2})^L$, 
so this factor is never included in the numerical result.
The decomposition produces 384 subsectors.

\begin{table}[htb]
\begin{center}
\begin{tabular}{|l||l|l|}
\hline
(s,t,u)&  (-1,-1,-1) & (-1,-2,-3)\\
\hline
$P_{-4}$ &  1.75006$\pm 1.3\times 10^{-4}$&0.41670$\pm 1.1\times  10^{-4}$\\
$P_{-3}$ &  -2.99969$\pm$ 0.00055 &-0.9313 $\pm$0.00067 \\
$P_{-2}$ &  -22.821 $\pm$ 0.003& -5.8599 $\pm$ 0.0035\\
$P_{-1}$ & 113.629 $\pm$0.013 & 42.79 $\pm$0.02 \\
$P_0$ & -395.27 $\pm$ 0.05 & -162.73$\pm$0.09\\ 
\hline
\end{tabular}
\end{center}
\caption{Numerical results for the points $(s,t,u)= (-1,-1,-1)$ and $(-1,-2,-3)$ of the massless non-planar double box.
\label{table:np7}}
\end{table}

The result for the graph called {\it NPbox} at the numerical point called {\it point} 
in the input file 
will be written to the file {\tt NPbox\_[point]full.res} in the subdirectory {\tt 2loop/NPbox}, 
where {\tt 2loop} is a subdirectory which has been created by the program, using  the 
directory name the user has specified in the first entry of {\tt paramNPbox.input}. 
By default, a subdirectory with the name of the graph is created, but the user can also 
specify a completely different directory (e.g. {\tt scratch}) where the results will be written to
(second entry in {\tt paramNPbox.input}).

More information about the decomposition is given in the file {\tt NPboxOUT.info}.
Information about the numerical integration is contained in the files \\{\tt [point]intfile.log} in 
the subdirectories {\tt graph/polestructure/epstothe[i]}, 
where ``polestructure" is of the form e.g. {\tt 2l0h0}, 
denoting 2 logarithmic poles and 0 linear, 0 higher poles.

It should be emphasized that in {\tt param.input}, the numbers for the Mandelstam 
invariants should be defined 
as the {\it Euclidean} values,  so the values for $s,t,u,p_i^2$ should always be 
negative in {\tt param.input}. 
Note also that the condition $s+t+u=0$ cannot be fulfilled numerically in the Euclidean region, 
so it should not be used in {\tt onshell=\{\ldots\}} in the template file 
to eliminate $u$ from the function ${\cal F}$ in the case of non-planar 
box graphs. 

\subsubsection{Example 2: Planar two-loop ladder diagram with massive on-shell legs}
\label{sec:QEDbox}
\begin{figure}
\begin{center}
\includegraphics[width=4.5cm]{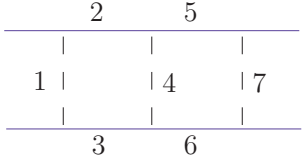}
\end{center}
\caption{Blue (solid) lines denote massive particles.}
\label{fig:qed}
\end{figure}

The purpose of this example is to show how to deal with diagrams where 
the decomposition could run into an infinite recursion if the default strategy is applied.
The rungs of the ladder are massless particles (e.g. photons), while the 
remaining lines are massive on-shell particles, depicted by solid (blue) lines in Fig.\,\ref{fig:qed}.
To run this example, execute the command {\it ../launch -p paramQED.input -t templateQED.m} from the {\tt loop/demos}
directory.
Only the primary sectors number one and seven are at risk of running into infinite recursion, 
therefore they are listed in the third-last item of {\tt paramQED.input} as the ones 
to be decomposed by a different strategy.
The results for the numerical point called {\it point} 
will be written to the file {\tt QED\_[point]full.res} in the subdirectory {\tt 2loop/QED}.
Numerical results for some sample points  are given in Table\,\ref{table:qed}. 
The kinematic points are defined by the mass $m$ and the Mandelstam variables 
$s=(p_1+p_2)^2, t=(p_2+p_3)^2$. We extracted a prefactor of $\Gamma(1+\eps)^2$.

\begin{table}[htb]
\begin{center}
\begin{tabular}{|l||l|l|}
\hline
(s,t,\,m)& (-0.2,-0.3,1) &(-3/2,-4/3,1/5)\\
\hline
$P_{-2}$ & -1.56161$\pm$ 1.33$\times 10^{-4}$ &-2.1817 $\pm$ 0.0003 \\
$P_{-1}$ &  -5.3373 $\pm$0.0018  &-1.4701 $\pm$0.0026\\
$P_0$ & 1.419 $\pm$0.025 & 30.191$\pm$ 0.014\\
$P_1$&  62.46 $\pm$ 0.18 &140.73$\pm$0.057\\
$P_2$ & 284.76 $\pm$ 0.87& 450.67$\pm$0.19\\ 
\hline
\end{tabular}
\end{center}
\caption{Numerical results up to order $\eps^2$ for the points $(s,t,m)=(-0.2,-0.3,1)$ and $(-3/2,-4/3,1/5)$ of the two-loop
ladder diagram shown in Fig.\,\ref{fig:qed}. An overall factor of $\Gamma(1+\eps)^2$ is 
not included in the numerical result.
\label{table:qed}}
\end{table}

\subsubsection{Example 3: Non-planar two-loop diagrams with two massive on-shell legs}

\begin{figure}
\begin{center}
\unitlength=1mm
\begin{picture}(100,50)
\put(-5,5){\includegraphics[width=6.cm]{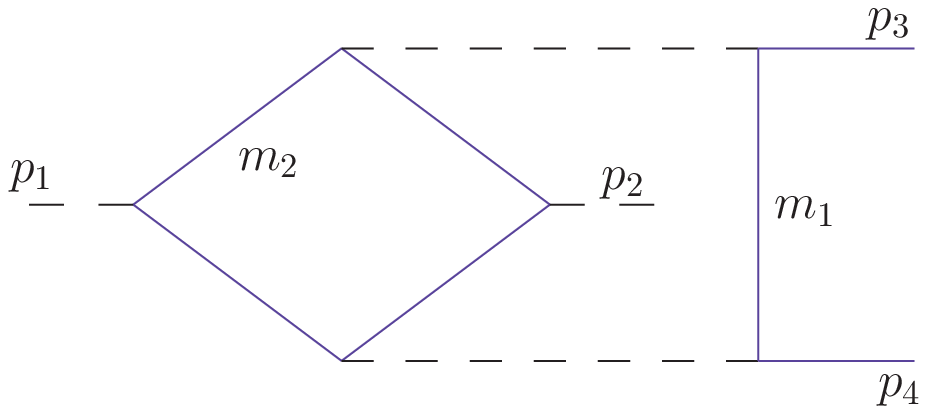}}
\put(20,0){ggtt1}
\put(55,5){\includegraphics[width=6.cm]{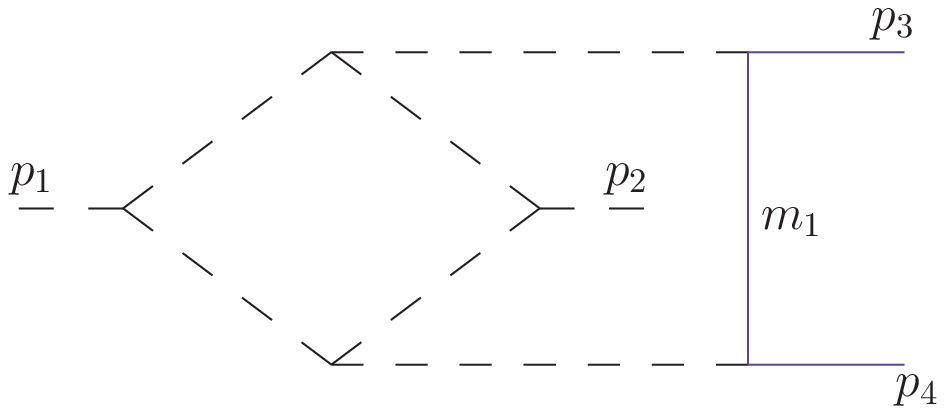}}
\put(80,0){ggtt2}
\end{picture}
\end{center}
\caption{Non-planar graphs occurring in the calculation of $gg\to t\bar{t}$ at NNLO.
Blue (solid) lines denote massive particles.}
\label{fig:ggtt}
\end{figure}

This example gives results for two non-planar graphs occurring in the calculation of 
$gg\to t\bar{t}$ at NNLO, shown in Fig.~\ref{fig:ggtt}. 
The analytic results for these graphs are not yet available.
Numerical results at Euclidean points
can be produced by choosing numerical values for the invariants $s,t,u,m^2$ 
in {\tt paramggtt1.input} respectively {\tt paramggtt2.input}
and then executing the command 
{\it ../launch -p paramggtt1.input -t templateggtt1.m} in the {\tt loop/demos} directory, analogously for {\it ggtt2}.
Results for two sample points are shown in Table \ref{table:ggtt}. 

\label{sec:ggtt}

\begin{table}[htb]
\begin{center}
\begin{tabular}{|l||l|l|}
\hline
\hline
\multicolumn{3}{|c|}{ggtt1}\\  
\hline
\hline
$(s,t,u,m_1^2,m_2^2)$& (-0.5,-0.4,-0.1,0.17,0.17) & (-1.5,-0.3,-0.2,3,1)\\
\hline
 $P_{0}$ & -38.0797$\pm$0.0027 &-0.19904$\pm 1.5\times 10^{-5}$ \\
 $P_{1}$ & -263.22$\pm$ 0.015& -0.71466$\pm 6 \times 10^{-5}$ \\
 $P_{2}$ &  -936.86$\pm$ 0.06& -1.45505$\pm$ 0.0002\\
\hline
\hline
\multicolumn{3}{|c|}{ggtt2}\\  
\hline
\hline
$(s,t,u,m_1^2,m_2^2)$& (-0.5,-0.4,-0.1,0.17,0)  &(-1.5,-0.3,-0.2,3,0)\\
\hline
$P_{-4}$ &  -10.9159  $\pm$ 0.0006 & -0.13678$\pm 1.46\times 10^{-5}$ \\
$P_{-3}$ &-43.5213   $\pm$ 0.0075 &-0.2087 $\pm$0.00024\\
$P_{-2}$ &  165.384  $\pm$ 0.048 &  3.3417 $\pm$0.0014\\
$P_{-1}$ & 20.842$\pm$0.268  &-6.593$\pm$0.007 \\
$P_0$ & 2117.5  $\pm$ 1. 57 & 20.42$\pm$0.04  \\ 
\hline
\end{tabular}\end{center}
\caption{Numerical results for the diagrams
shown in  Fig.~\ref{fig:ggtt}. The finite diagram ggtt1 has been calculated up to order $\eps^2$.
An overall factor of $\Gamma(1+\eps)^2$ is extracted.\label{table:ggtt}
}
\end{table}

\subsubsection{Example 4:  A rank one tensor two-loop box}

In order to demonstrate how to run the program for integrals with non-trivial numerators, 
we give the example of a  rank one planar massless on-shell two-loop box, where we contract 
one loop momentum in the numerator by $2\,p_3^{\mu}$.
\bea
G=\int \frac{d^Dk\,d^Dl}{(i\pi^{\frac{D}{2}})^2} 
\frac{ 2\,p_3\cdot k}{k^2 (k-p_1)^2 (k+p_2)^2 (k-l)^2 (l-p_1)^2 (l+p_2)^2  (l+p_2+p_3)^2}
\;,\nn\\
\label{eq:dbr1}
\eea
where we omitted the $i\delta$ terms in the propagators.
The result for the kinematic sample point $(s,t,u)=(-3,-2,5)$ is shown in Table\,\ref{table:DBr1}.
\begin{table}[htb]
\begin{center}
\begin{tabular}{|l||l|}
\hline
(s,t,u)& (-3,-2,5) \\
\hline
$P_{-4}$ & -0.319449 $\pm 1.7\times 10^{-5}$ \\
$P_{-3}$ &0.46536 $\pm 8\times 10^{-5}$\\
$P_{-2}$ & 0.5848 $\pm$0.0004 \\
$P_{-1}$ &  -3.3437 $\pm$ 0.0013\\
$P_0$ & -1.6991$\pm$  0.0035\\ 
\hline
\end{tabular}\end{center}
\caption{Numerical results for the point $(s,t,u)= (-3,-2,5)$ of the rank one two-loop
ladder diagram given by eq.~(\ref{eq:dbr1}). 
An overall factor of $\Gamma(1+\eps)^2$ has been extracted.
\label{table:DBr1}
}
\end{table}
Note that in this example, we used a positive value for the Mandelstam invariant $u$,
which seems to contradict the requirement to have only Euclidean values for the invariants.
However, in this case we can do this because the function ${\cal F}$ does not depend 
on $u$ at all. The numerator does depend on $u$, but as a numerator which is not positive definite 
does not spoil the numerical convergence, we can as well choose a numerical value 
for $u$ such that the relation $s+t+u=0$ is fulfilled. This has the advantage that it allows us to 
use the latter relation to simplify the numerator.

\subsubsection{Example 5:  A three-loop vertex diagram with $\eps$-dependent propagator powers}

This example shows how to calculate diagrams with propagator powers different from one.
The results for the graph $A_{6,1}$ (notation of Ref.\,\cite{Gehrmann:2006wg}), 
given in Table \ref{table:A61}, can be produced by running 
{\it ../launch -p paramA61.input -t templateA61.m } from the {\tt loop/demos} directory.

\begin{figure}
\begin{center}
\includegraphics[width=4.cm]{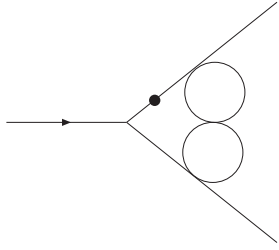}
\end{center}
\caption{The three-loop vertex diagram $A_{6,1}$ with the dotted propagator raised to the power $1+\eps$.\label{fig:a61}}
\end{figure}

\begin{table}[htb]
\begin{center}
\begin{tabular}{|l|l|l|l|l|l|}
\hline
  $P_{-3}$ &$P_{-2}$ & $P_{-1}$ & $P_0$& $P_1$& $P_2$ \\ \hline
   0.16666  &  1.8334 & 18.123 & 125.32 & 889.96& 5325.3 \\ 
\hline
\end{tabular}
\end{center}
\caption{Numerical results for the diagram shown in Fig.~\ref{fig:a61} with 
the dotted propagator 
raised to the power $1+\eps$. The errors are below one percent. 
\label{table:A61}}
\end{table}
The analytical result for this diagram with general propagator powers is given in Ref.~\cite{Gehrmann:2006wg} 
and is also given in the file {\tt 3loop/A61/A61analytic.m} to allow comparisons between 
analytical and numerical results for arbitrary propagator powers.

\subsection{More general polynomial functions}

The examples described below can be found in the subdirectory {\tt general/demos}.

\subsubsection{Example 6:  Hypergeometric functions}

As an example for ``general" polynomial functions, 
we consider the  hypergeometric functions 
$_pF_{p-1}(a_1,\ldots,a_p;b_1,\ldots,b_{p-1};\beta)$, 
using the  integral representation recursively: 
\bea
&&_pF_{p-1}(a_1,\ldots,a_p;b_1,\ldots,b_{p-1};\beta)=\frac{\Gamma(b_{p-1})}{\Gamma(a_p)\Gamma(b_{p-1}-a_p)}\label{pFq}\\
&&\int_0^1 dz\,(1 - z)^{-1 -
a_p + b_{p-1}}\,z^{-1 + a_p}\,_{p-1}F_{p-2}(a_1,\ldots,a_{p-1};b_1,\ldots,b_{p-2};\beta)\;,\nn\\
&&\nn\\
&&_2F_{1}(a_1,a_2;b_1;\beta)=
\frac{\Gamma(b_{1})}{\Gamma(a_2)\Gamma(b_{1}-a_2)}\int_0^1 dz\,(1 - z)^{-1 -
a_2 + b_{1}}\,z^{-1 + a_2}\,(1 - \beta \,z)^{-a_1}\;.\nn
\eea
Considering $_5F_4(a_1,\ldots,a_5;b_1,\ldots,b_{4};\beta )$ with the values 
$a_1=\epsilon,a_2=-\epsilon,a_3=-3\epsilon,a_4=-5\epsilon,a_5=-7\epsilon$,
$b_1=2\epsilon,b_2=4\epsilon,b_3=6\epsilon,b_4=8\epsilon, \beta=0.5$
we obtain the results shown in Table \ref{table:5F4}.
The ``analytic result" has been obtained using HypExp\,\cite{Huber:2005yg,Huber:2007dx}.

\begin{table}[htb]
\begin{center}
\begin{tabular}{|c|c|c|c|}
\hline
$\epsilon$ order & analytic result & numerical result & time taken (secs)\\
\hline
$\epsilon^0$ & 1 &1.0000002 $\pm 4\times 10^{-7}$ & 2\\
$\epsilon^1$ & 0.189532& 0.189596$\pm$0.00036 & 21 \\
$\epsilon^2$ & -2.299043&-2.306$\pm$0.011 & 124 \\
$\epsilon^3$ & 55.46902 &55.61 $\pm$0.39 & 248 \\
$\epsilon^4$ & -1014.39 & -1018.4$\pm$5.9 & 429\\
\hline
\end{tabular}
\caption{Results for the hypergeometric function $_5F_4(\eps,-\eps,-3\eps,-5\eps,-7\eps;2\eps,4\eps,6\eps,8\eps;\beta )$  at $\beta=0.5$.
The timings in the last column are the ones for the numerical integration. The time taken for decomposition,
subtraction and $\eps$-expansion was 11 seconds.
\label{table:5F4}}
\end{center}
\end{table}
This result can be produced by typing 
{\it ./launch -d demos -p param5F4.input -t template5F4.m} 
in the subdirectory {\tt general}, 
or by typing {\it ../launch  -p param5F4.input -t template5F4.m} 
in the subdirectory {\tt general/demos}.

The program can also deal with functions containing half integer exponents. 
Table \ref{table:4F3} shows results for $_4F_3$ with arguments 
$a_1=-4\eps,
a_2=-1/2-\eps,
a_3=-3/2-2\eps,
a_4=1/2-3\eps,
b_1=-1/2+2\eps,
b_2=-1/2+4\eps,
b_3=1/2+6\eps$.
These results can be produced by the command 
{\it ../launch  -p param4F3.input -t template4F3.m} 
in the subdirectory {\tt general/demos}.

\begin{table}[htb]
\begin{center}
\begin{tabular}{|c|c|c|c|}
\hline
$\epsilon$ order & analytic result  & numerical result & time taken (seconds)\\
\hline
$\epsilon^0$ &1 &0.999997 $\pm 1.7\times 10^{-5}$&1.6\\
$\epsilon^1$ &-4.27969 &-4.2810 $\pm$ 0.0055& 54\\
$\epsilon^2$ &-26.6976 &-26.625 $\pm$0.121 &90\\
\hline
\end{tabular}
\caption{Results for the hypergeometric function $_4F_3(-4\eps,-1/2-\eps,-3/2-2\eps,1/2-3\eps;-1/2+2\eps,-1/2+4\eps,1/2+6\eps;\beta )$  at $\beta=0.5$.
\label{table:4F3}}
\end{center}
\end{table}

\subsubsection{Example 7:  Phase space integrals}

Sector decomposition can be useful for the calculation of phase space integrals 
where infrared divergences are regulated dimensionally.
This is  particularly the case for double real radiation 
occurring in NNLO calculations involving massive particles, where analytic methods show their limitations.

Here we give  examples of  $2\to 3$ phase space integrals, which should be considered as part of a 
$2\to n$ phase space written in factorised form. 
We choose particles 3 and 4 to be massless, while $p_5$ is the momentum of a massive state, 
either a single particle or a pseudo-state formed by $n$
additional momenta $\tilde{p}_i$ in the final state, i.e. $p_5=\sum_{i=5}^n \tilde{p}_i$. 
After all integrations have been mapped to the unit interval, we have integrals of the form
\bea
\int d\Phi_{3}&=&C_\epsilon \,
 \int \prod_{i=1}^4 d x_i  
 \,\left[x_1(1-x_1)x_2(1-x_2)\right]^{\frac{D-4}{2}}\,[x_3 \,(1-x_3)]^{D-3}\nn\\
&&
[x_4\,(1-x_4)]^{\frac{D-5}{2}}\,
\left[1-\beta\,x_3\,(1-x_2)\right]^{2-D}\;,\label{PSa}\\
\beta&=&1-\frac{m^2}{s}\;,\; C_\epsilon =
\frac{1}{(2\pi)^{2D-3}}d\Omega_{D-3}d\Omega_{D-4}\,s^{D-3} 2^{D-8}\beta^{2D-5}
\;.
\label{ceps}
\eea
The derivation is given in the appendix, section \ref{appendix:ps}.

The invariants in this parametrisation are given by
\bea
s_{13}&=&-s\beta\,x_3\,(1-x_1)\nn\\
s_{23}&=&-s\beta\,x_3\,x_1\nn\\
s_{34}&=&\beta\,K\,x_3\,(1-x_2)\;,\;K=\frac{s\beta\,(1-x_3)}{1-\beta\,x_3\,(1-x_2)}\nn\\
s_{35}&=&s\,\frac{1-\beta(1-x_2x_3)}{1-\beta\,x_3(1-x_2)}\nn\\
s_{14}&=&-K\,\left\{ t^- + x_4 \,(t^+-t^-)\right\}=-K\,\,\tilde{s}_{14}\nn\\
s_{24}&=&-K\,\left\{ u^+ - x_4 \,(u^+-u^-)\right\}=-K\,\,\tilde{s}_{24}\;,\nn
\eea
where 
\bea
t^\pm&=&\left(\sqrt{x_1(1-x_2)}\pm\sqrt{x_2(1-x_1)} \right)^2\label{tpm}\\
u^\pm&=&\left(\sqrt{(1-x_1)(1-x_2)}\pm\sqrt{x_1x_2}  \right)^2\nn\;.
\eea

We would like to point out that for the examples below, more convenient parametrisations, 
i.e. parametrisations where the variables in the denominator factorise, 
and/or reflect symmetries of the squared matrix element, certainly do exist.
However, the purpose of the examples is to illustrate that the code can deal with 
denominators which are amongst the most complicated ones which do occur in 
NNLO real radiation involving two (unresolved) massless particles in the final state,
where they cannot always be ``rotated away" by suitable transformations.
A hybrid approach combining sector decomposition with convenient 
parametrisations/transformations is certainly the method of choice for real radiation at 
NNLO. The program can be used to evaluate the integrals occurring in such an approach. 

\subsubsection*{Three massless particles in the final state}

\begin{figure}
\begin{center}
\unitlength=1mm
\begin{picture}(70,20)
\put(-5,5){\includegraphics[width=6.cm]{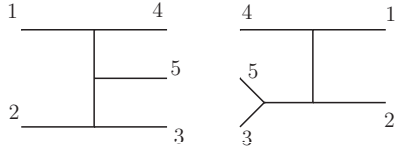}}
\end{picture}
\end{center}
\caption{Interference of diagrams   leading to factors of  $s_{35}s_{23}$ in the denominator. \label{fig:s23s35}}
\end{figure}

We first consider a case where $p_5$ is a massless particle, i.e. the limit $\beta\to 1$ in 
eq.\,(\ref{PSa}).
If we combine the phase space with the toy matrix element
$1/(s_{35}s_{23})$, we have singularities at  $x_1=0$, $x_2=0$ and $x_3=0$. 
Such denominators come e.g. from the interference of diagrams as shown in Fig. \ref{fig:s23s35}.

\bea
\int d\Phi_{3}\,\frac{s^2}{s_{35}s_{23}}&=&C_\eps\,
\int_0^1\prod_{i=1}^4 d x_i  
 \,\left[(1-x_1)(1-x_2)\right]^{\frac{D-4}{2}}\,[x_1\,x_2]^{\frac{D-6}{2}}\,x_3^{D-5} \,(1-x_3)^{D-3}\nn\\
&&
[x_4\,(1-x_4)]^{\frac{D-5}{2}}\,
\left[1-\,x_3\,(1-x_2)\right]^{3-D}\;.\label{eq:PS3523}
\eea
The term $[1-\,x_3\,(1-x_2)]^{3-D}$ goes to zero for $x_3\to 1, x_2\to 0$.
Although $x_3=1$ does not lead to a singularity in the above example, 
for numerical stability reasons, and having in mind 
the presence of more complicated matrix elements than our toy example,  it 
is preferable to transform this factor to an expression which is finite in the above limits. 
Splitting the $x_3$ integration at 1/2 and then doing sector decomposition achieves this goal.
The program will do this automatically if the template file contains {\tt splitlist=\{3\}}, 
to tell the program that the integration over $x_3$ should be split at 1/2. 
Of course the singularities at $x_i=0$ will also be extracted automatically.

Using the command  
{\it ../launch  -p params23s35.input -t templates23s35.m} 
in the subdirectory {\tt general/demos}, sector decomposition 
leads to the result given in Table \ref{table:ps35}.

\begin{table}[htb]
\begin{center}
\begin{tabular}{|l|l|l|l|}
\hline
$P_{-3}$ &$P_{-2}$ &  $P_{-1}$ & $P_0$\\ \hline
 -1.5705$\pm$  0.0005& -4.3530 $\pm$ 0.0025& 1.712$\pm$ 0.005 & 31.040$\pm$ 0.014\\ 
\hline
\end{tabular}\end{center}
\caption{Results for the integral given by eq.\,(\ref{eq:PS3523}). 
The factor $C_\eps$ is not included in the numerical result.
\label{table:ps35}}
\end{table}

\subsubsection*{Two massless and one massive particles in the final state}

The example in this subsection illustrates  the program option to exclude 
certain parts of the integrand from the decomposition, even though 
they can become zero  at certain values of the integration parameters. 
This can be useful if a particular term is known not to lead to a singularity.
Note that terms with powers $\geq 0$ are excluded from the decomposition by default.

As an example we pick an integral over $s_{14}$,  where the line singularity 
has been remapped already (see appendix, section  \ref{appendix:ps}).
\bea
&&\int d\Phi_{3}\,\frac{s\,\beta}{s_{14}}=
2\,C_\epsilon \,
 \int_0^1 dx_1 dx_2 dx_3 dx_4 [x_4\,(1-x_4)]^{\frac{D-5}{2}}\,x_3^{D-3} \,(1-x_3)^{D-4}\label{eq:pss14}\\
&&\left[1-\beta\,x_3\,(1-x_1+x_1x_2)\right]^{3-D}
 x_1^{D-4}\,x_2^{D-5}\,\left[(1-x_1)(1-x_2)(1-x_1+x_1x_2)\right]^{\frac{D-4}{2}}\nn\\
&& \left[ (\sqrt{(1-x_1)(1-x_2)}-\sqrt{1-x_1+x_1x_2})^2 +4\,x_4\,\sqrt{(1-x_1)(1-x_2)(1-x_1+x_1x_2)}\right]^{4-D}\;.\nn
\eea

Choosing the option ``n" for ``no decomposition" in the definition of the integrand 
for the term in square brackets $[\ldots]^{4-D}$
(see {\tt templates14.m}), there will be no decomposition in the variables 
$x_2,x_4$, although  this term vanishes in the limit $x_2,x_4\to 0$, but this limit 
does not lead to a singularity. 
The result, which can be produced by 
{\it ../launch  -p params14.input -t templates14.m} 
in the subdirectory {\tt general/demos}, 
 is given in Table \ref{table:ps14}.

\begin{table}[htb]
\begin{center}
\begin{tabular}{|l|l|}
\hline
  $P_{-1}$ & $P_0$\\ \hline
  -1.12635 $\pm$ 0.0003 & -8.771 $\pm$  0.003 \\ 
\hline
\end{tabular}\end{center}
\caption{Numerical result for the integral given by eq.\,(\ref{eq:pss14}) for $\beta$=0.75. 
The factor $2\,C_\epsilon$ is not included in the numerical result.
\label{table:ps14}}
\end{table}

\section{Conclusions and outlook}

We have presented a program for the numerical 
evaluation of multi-loop integrals in Euclidean space, 
as well as  the evaluation of
more general parameter integrals in the context of dimensional regularisation.
Singularities which lead to poles in 1/$\epsilon$ are extracted automatically using 
iterated sector decomposition.  The program then produces finite functions 
forming  the coefficients of 
a Laurent series in $\epsilon$, which are evaluated  numerically by Monte Carlo integration.  

In the case of loop integrals, the program can deal with arbitrary tensor  integrals, 
where the corresponding numerator function is constructed automatically.
Non-standard propagator powers, e.g. powers depending on $\eps$, are also supported.
In the case of general polynomial functions, the program also can deal with
cases where the integration parameters or the polynomials present in the integrand
are raised to  half-integer powers. 
Constants can be left symbolic at the sector decomposition stage; their values 
can be specified later  at the numerical integration stage, such that the decomposition 
has to be done only once and for all while the numerical values for the constants 
still can be changed.

The code is publicly available at 
{\tt http://projects.hepforge.org/secdec/} 
and comes with various examples and detailed documentation.
Different choices of numerical integration packages are possible, i.e. 
the Monte Carlo program Bases~\cite{Kawabata:1995th} and the ones 
from the Cuba library given in \cite{Hahn:2004fe}.
For a future version, we plan to  include alternative tools  to generate 
optimised functions, e.g. the one described in \cite{Reiter:2009ts}.
A  version which includes  non-linear transformations in an automated way, 
combined with the extension to integrands containing e.g. physical thresholds,  
requiring complex contour integration,  
is also planned.

\section{Acknowledgements}

We would like to thank Daniel Ma{\^i}tre for  useful comments 
on  the program and the manuscript.
We also are grateful to P.-F. Monni, M.~Zentile and G.~Welsh for testing the program, and
C.~Studerus for conversation on the master integrals for $t\bar{t}$ production.
This research was supported by the British Science and Technology Facilities Council (STFC).

\section{Appendix}

\subsection{Timings for a 4-loop two-point diagram}

\begin{figure}[htb]
\begin{center}
\unitlength=1mm
\begin{picture}(40,25)
\put(-5,0){\includegraphics[width=4.cm]{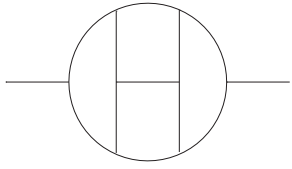}}
\end{picture}
\end{center}
\caption{A four-loop two-point master integral \label{fig:4loop}}
\end{figure}

In order to give an idea about the timings for a complicated example which 
we ran on several processors, 
we give the timings for the four-loop graph shown 
in Fig.\,\ref{fig:4loop}.
The coefficient 
of each pole order is composed of a number of functions which can be integrated 
individually, such that the time taken for the longest job equals the total time for a given pole order, 
provided that the contributing functions are integrated in parallel. 
The files to run the integration of these functions in parallel 
 are created by the program automatically. 
 The  number  of integrations to run individually depends on the size 
 and number of  the regular 
 subsector  functions contributing at each pole order: 
  these functions are summed until their sum reaches about one Megabyte, 
  and then integrated individually, preferably in parallel.

\begin{table}[htb]
\begin{center}
\begin{tabular}{|c|c|c|c|}
\hline
Stage&Time for longest job (secs)& nb of functions &Total time (secs)\\
\hline
Sector Decomposition & 615 & & 2160\\
Subtraction \& $\epsilon$-expansion & 809 && 3767\\
Numerical integration $\epsilon^{-1}$ & 156 & 5 & 508\\
Numerical integration $\epsilon^{0}$ & 422 & 28 & 4720\\
Numerical integration $\epsilon^{1}$ & 492 & 28 &  5946\\
Numerical integration $\epsilon^{2}$ & 2172 & 29 & 8123\\
\hline
\multicolumn{3}{c}{}
\end{tabular}
\end{center}
\caption{Timings for the diagram shown in Fig. \ref{fig:4loop}.  
The time taken for the longest job equals the total time for a given pole order 
if the contributing functions are integrated in parallel. The number of 
sampling points was 500000 for each pole order. The last column shows the
 timings which would result from a calculation in series.
\label{table:4loop}}
\end{table}

The timings are listed in Table   \ref{table:4loop}.
For information we also give 
the timings which would result from a serial calculation in the  third column of 
Table \ref{table:4loop}.
The results are shown in Table \ref{table:4loopresult}.

\begin{table}[htb]
\begin{center}
\begin{tabular}{|c|c|c|}
\hline
Order & Analytical result & Numerical result\\
\hline
$\epsilon^{-1}$ & -10.3692776 & -10.371$\pm$0.002\\
$\epsilon^{0}$ & -70.99081719 & -71.002$\pm$0.013\\
$\epsilon^{1}$ & -21.663005 & -21.65$\pm$0.12\\
$\epsilon^{2}$ & Unknown & 2833.79$\pm$0.92\\
\hline
\end{tabular}
\end{center}
\caption{``Analytical" and numerical results for the diagram shown in Fig. \ref{fig:4loop}.  
The analytical result can be found in \cite{Baikov:2010hf}.
\label{table:4loopresult}}
\end{table}

In this calculation, the symmetry of the problem was used, so only four primary
sectors were evaluated. The corresponding multiplicities of the primary sectors 
are taken into account automatically, provided they are specified 
in {\tt param.input}.


\subsection{Another representation of the non-planar two-loop box}
\label{appendix:xbox}

Here we derive a representation of the non-planar two-loop box where one 
integration parameter factorises naturally, such that it can be integrated out analytically, 
leaving a representation which  can be evaluated in a completely 
automated way by the  routines in {\tt SecDec/general}, 
the evaluation being considerably faster than the one of example \ref{example:npbox}.
This procedure is not limited to our particular example, but requires an analytical step
of introducing a convenient parametrisation.

\begin{figure}[htb]
\begin{center}
\unitlength=1mm
\begin{picture}(40,25)
\put(-5,0){\includegraphics[width=4.cm]{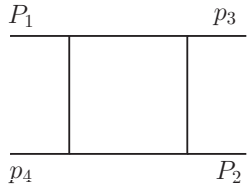}}
\end{picture}
\end{center}
\caption{The ``inner" box as part of the non-planar two-loop box shown in 
Fig.\,\ref{fig:np7}.\label{fig:xbox}}
\end{figure}

The expression for the non-planar two-loop box shown in 
Fig.\,\ref{fig:np7} is given by 
\bea
G_{NP}=\int \frac{d^Dk\,d^Dl}{(i\pi^{\frac{D}{2}})^2} 
\frac{1}{k^2 (k+p_2)^2(k-p_1)^2  (k-l)^2 (l+p_2)^2 (k-l+p_4)^2 (l+p_2+p_3)^2}\;.\nn\\
\eea
Considering first the integration over the loop momentum $l$ only, we have a one-loop 
box as shown in Fig.\,\ref{fig:xbox} with $P_1=p_1-k, P_2=p_2+k$.
Feynman parametrisation for this one-loop subgraph leads to
\bea
I_1&=&\int \frac{d^Dl}{i\pi^{\frac{D}{2}}} 
\frac{1}{(k-l)^2 (l+p_2)^2 (k-l+p_4)^2 (l+p_2+p_3)^2}\nn\\
&=&\Gamma(2+2\eps) \int\prod\limits_{i=1}^{4} dx_i\,\delta(1-\sum_{j=1}^4 x_j)\,{\cal F}(\vec{x},k)^{-2-\eps}\label{i1}\\
{\cal F}(\vec{x},k)&=&-(p_1+p_4-k)^2\,x_1x_3-(p_1+p_3-k)^2\, x_2x_4-(k+p_2)^2\,x_1x_2-(p_1-k)^2\,x_3x_4\;.\nn
\eea
Now we substitute 
\bea
x_1=t_2\,(1-t_3)\,,\,x_2=t_1t_3\,,\,x_3=(1-t_1)\,t_3
\eea
and integrate out the $\delta$-constraint to obtain 
\bea
I_1
&=&\Gamma(2+\eps) \int_0^1 dt_1dt_2dt_3 \,[t_3\,(1-t_3)]^{-1-\eps}\nn\\
&&\left[
-(p_1+p_4-k)^2\,t_2\bar{t}_1-(p_1+p_3-k)^2\,t_1\bar{t}_2 -(k+p_2)^2\,t_1t_2-(p_1-k)^2\,\bar{t}_1\bar{t}_2\right]^{-2-\eps}\nn\\
&=&\frac{2\,\Gamma(2+\eps)\Gamma(-\eps)\Gamma(1-\eps)}{\Gamma(1-2\eps)}
\int_0^1 dt_1dt_2\label{t3eli}\\
&&\left[
-(p_1+p_4-k)^2\,t_2\bar{t}_1-(p_1+p_3-k)^2\,t_1\bar{t}_2 -(k+p_2)^2\,t_1t_2-(p_1-k)^2\,\bar{t}_1\bar{t}_2\right]^{-2-\eps}\;,\nn
\eea
where we used the shorthand notation $\bar{t}_i=1-t_i$.
Now we combine the above expression with the remaining propagators, 
treating the expression in square brackets in eq.\,(\ref{t3eli}) as a fourth propagator
with power $2+\eps$. One can use the {\tt SecDec} code to calculate the resulting 
function integrand function ${\cal F}_2$, although this is an easy calculation to do by hand.
One obtains
\bea
G_{NP}&=&\frac{-2\,\Gamma(3+2\eps)\Gamma(-\eps)\Gamma(1-\eps)}{\Gamma(1-2\eps)}
\int_0^1 dt_1dt_2  \int\prod\limits_{i=1}^{4} dz_i\,z_4^{1+\eps}\delta(1-\sum_{j=1}^4 z_j)\,{\cal F}_2(\vec{z})^{-3-2\eps}\nn\\
{\cal F}_2(\vec{z})&=&-s_{12}\,z_2z_3-T\,z_1z_4-P_3^2\,z_2z_4-P_4^2\,z_3z_4\;,\nn\\
&&\mbox{where}\nn\\
T&=&s_{23}\,t_2(1-t_1)+s_{13}\,t_1(1-t_2)\;,\;s_{ij}=(p_i+p_j)^2\;,\nn\\
P_3^2&=&s_{12}\,(1-t_1)(1-t_2)\;,\;P_4^2=s_{12}\,t_1t_2\;.
\eea
With the substitutions 
\bea
z_4=t_3t_4\,,\,z_3=t_3\,(1-t_4)\,,\,z_2=(1-t_3)\,t_5\,,\,z_1=(1-t_3)\,(1-t_5)
\eea
we finally obtain 
\bea
G_{NP}&=&\frac{-2\,\Gamma(3+2\eps)\Gamma(-\eps)\Gamma(1-\eps)}{\Gamma(1-2\eps)}
\int_0^1 dt_1dt_2  dt_3dt_4dt_5\,t_3^{-1-\eps}(1-t_3)\,t_4^{1+\eps}\,\left[\,{\cal F}_2(\vec{t})\,\right]^{-3-2\eps}\nn\\
{\cal F}_2(\vec{t})&=&-s_{12}\,\bar{t}_3\bar{t}_4t_5-T\,\bar{t}_3t_4\bar{t}_5-
P_3^2\,\bar{t}_3t_4t_5-P_4^2\,t_3t_4\bar{t}_4\;,\;\,
\bar{t}_i=1-t_i\;.\nn
\eea
In this form the integrand can be fed into the sector decomposition routine for ``general integrands".
The corresponding template file {\tt templatexbox.m} can be found in {\tt SecDec/general/demos}. 

\subsection{Phase space parametrisation of a 2 $\to$ 3 phase space}
\label{appendix:ps}

The $D-$dimensional phase space for $p_1+p_2\to p_3+p_4+p_5$
is given by 
\bea\label{Eq:PSfull}
d\Phi_{3} &=&  
\Bigl[ \prod\limits_{j=3}^{5} 
\frac{d^D p_j}{(2\pi)^{D-1}}  \Bigr] \,
\delta^+(p_3^2-m_3^2)\delta^+(p_4^2-m_4^2)
\delta^+(p_5^2-m_5^2)
(2\pi)^D\delta^{(D)}\Bigl(p_1+p_2-\sum\limits_{j=3}^{5} p_j \Bigr)\nn\;.
\eea
Using 
\begin{eqnarray*}
\int d^Dp_j\,\delta^+(p_j^2-m_j^2)
&=&\int d^{D-1}\vec{p}_j\,dE_j\,\delta(E_j^2-\vec{p}_j^{\,2}-m_j^2)\,\theta(E_j)\nn\\
&=&\frac{1}{2}\int
d E_j \,d\Omega_{D-2}^{(j)}\, |\vec{p}_j|^{D-3}  \Big|_{|\vec{p}_j|=\sqrt{E_j^{\,2}-m_j^2}}\;,\\
\int d\Omega_{D-2}&=&\frac{2\pi^{\frac{D-1}{2}}}{\Gamma(\frac{D-1}{2})}\;,\nn
\end{eqnarray*}
and eliminating $p_5$ by momentum conservation, 
one obtains
\bea
 \int d\Phi_{3}
 &=&\frac{1}{(2\pi)^{2D-3}}\frac{1}{4}\int 
d E_3\,d E_4\,\,d\Omega_{D-2}^{(3)}\, |\vec{p}_3|^{D-3} 
\,d\Omega_{D-2}^{(4)}\, |\vec{p}_4|^{D-3} \nn\\
&&\delta((p_1+p_2-p_3-p_4)^2-m_5^2)\;.
\eea
Having NNLO phase spaces in mind, let us assume that we have two massless particles
in the final state (which may become unresolved), so $m_3^2=m_4^2=0$, and $p_5$ 
is the momentum of a massive state, $m_5^2=m^2$, either a single particle or a pseudo-state formed by $n$
additional momenta $\tilde{p}_i$ in the final state, i.e. $p_5=\sum_{i=5}^n \tilde{p}_i$. 
In this case we can parametrise the momenta $p_1\ldots p_4$ as 
\bea
p_1 &=& \frac{\sqrt{s}}{2}\, (1,\vec 0^{(D-2)},1) \nn\\  
p_2 &=& \frac{\sqrt{s}}{2}\, (1,\vec 0^{(D-2)},-1) \nn\\
p_3 &=& E_3\, (1,\vec 0^{(D-4)},\sin\eta\sin\theta_1,\cos\eta\sin\theta_1,\cos\theta_1) \label{momenta}\\
&=&E_3\,(1,\vec 0^{(D-4)},\vec{n}_3)\nn\\
p_4 &=&E_4\, (1,\vec 0^{(D-4)},\vec{n}_4)\nn
\eea
where we choose $\vec{n}_4$ such that
\bea
\vec{n}_4&=&\left(\begin{array}{ccc}
1&0&0\\
0&\cos\theta_1&\sin\theta_1\\
0&-\sin\theta_1&\cos\theta_1
\end{array}\right)\left(\begin{array}{c}
\sin\phi\sin\theta_2\\
\cos\phi\sin\theta_2\\
\cos\theta_2
\end{array}\right)=\left(\begin{array}{c}
\sin\phi\sin\theta_2\\
\cos\theta_1\cos\phi\sin\theta_2+\sin\theta_1\cos\theta_2\\
\cos\theta_1\cos\theta_2-\sin\theta_1\cos\phi\sin\theta_2
\end{array}\right)\nn
\eea
Thus 
\bea 
d\Omega_{D-2}^{(3)}&=&d(\cos\theta_1)\,(1-\cos^2\theta_1)^{\frac{D-4}{2}}\,
d\Omega_{D-3}^{(\eta)}\nn\\
d\Omega_{D-2}^{(4)}&=& d(\cos\theta_2)\,(1-\cos^2\theta_2)^{\frac{D-4}{2}}\,
d\phi\,(\sin\phi)^{D-4}\,d\Omega_{D-4}\;.
\eea
Due to overall rotational invariance around the beam axis, 
we can integrate out the azimuthal angle $\eta$ and use $\eta=0$ in eq.~(\ref{momenta}), 
leading to $\vec{n}_3\cdot \vec{n}_4=\cos\theta_2$.  
We obtain
\bea
d\Phi_{3}&=&
\frac{1}{4}\frac{1}{(2\pi)^{2D-3}}d\Omega_{D-3}d\Omega_{D-4}\,
d(\cos\theta_1) \,d(\cos\theta_2)\, d(\cos\phi) \,d E_3\Theta(E_3)\,
d E_4\Theta(E_4)
\nn\\
&& \left(E_3 E_4\right)^{D-3}\left(\sin^2\theta_1\,\sin^2\theta_2\right)^{\frac{D-4}{2}}
(\sin^2\phi)^{\frac{D-5}{2}}\nn\\
&&\delta(m^2-(p_1+p_2-p_3-p_4)^2)\label{Eq:etheta}
\eea
$$
\delta(m^2-(p_1+p_2-p_3-p_4)^2)=\delta(s-m^2-2\sqrt{s}\,(E_3+E_4) +2\,E_3 E_4(1-\vec{n}_3\cdot \vec{n}_4))\;.$$
Now we have to choose which variable to eliminate using the $\delta$-constraint.
In our example,  we will eliminate  $E_4$, leading to
\be
E_4=\frac{s-m^2-2\sqrt{s}\,E_3}{2(\sqrt{s}-E_3\,(1-\vec{n}_3\cdot \vec{n}_4))}\;.
\ee
The constraint $\Theta(E_4)$ leads to 
\be
E_3^{\rm{max}}=\frac{s-m^2}{2\sqrt{s}}=\frac{\sqrt{s}}{2}\,\beta\;\mbox{ where } \beta=1-\frac{m^2}{s}\;.
\ee
We substitute 
\bea
E_3&=&\frac{\sqrt{s}}{2}\,\beta\,x_3 \Rightarrow E_4=\frac{\sqrt{s}}{2}\,\beta\,\frac{1-x_3}{1-\beta\,x_3(1-x_2)}\nn\\
\cos\theta_1&=&2\,x_1-1\;,\,
\cos\theta_2=2\,x_2-1\;,\,
\cos\phi=2\,x_4-1\nn
\eea
to obtain
\bea
d\Phi_{3}&=&
\frac{1}{(2\pi)^{2D-3}}d\Omega_{D-3}d\Omega_{D-4}\,s^{D-3} 2^{D-8}\beta^{2D-5}\nn\\
&& \prod_{i=1}^4 d x_i  
 \,\left[x_1(1-x_1)x_2(1-x_2)\right]^{\frac{D-4}{2}}\,[x_3 \,(1-x_3)]^{D-3}\nn\\
&&
[x_4\,(1-x_4)]^{\frac{D-5}{2}}\,
\left[1-\beta\,x_3\,(1-x_2)\right]^{2-D}\;.\label{appendix:psa}
\eea
The invariants in this parametrisation are given by
\bea
s_{13}&=&-s\beta\,x_3\,(1-x_1)\nn\\
s_{23}&=&-s\beta\,x_3\,x_1\nn\\
s_{34}&=&\beta\,K\,x_3\,(1-x_2)\;,\;K=\frac{s\beta\,(1-x_3)}{1-\beta\,x_3\,(1-x_2)}\nn\\
s_{35}&=&s\,\frac{1-\beta(1-x_2x_3)}{1-\beta\,x_3(1-x_2)}\nn\\
s_{14}&=&-K\,\left\{ t^- + x_4 \,(t^+-t^-)\right\}=-K\,\,\tilde{s}_{14}\nn\\
s_{24}&=&-K\,\left\{ u^+ - x_4 \,(u^+-u^-)\right\}=-K\,\,\tilde{s}_{24}\;,\nn
\eea
where 
\bea
t^\pm&=&\left(\sqrt{x_1(1-x_2)}\pm\sqrt{x_2(1-x_1)} \right)^2\label{tpmapp}\\
u^\pm&=&\left(\sqrt{(1-x_1)(1-x_2)}\pm\sqrt{x_1x_2}  \right)^2\nn\;.
\eea
Physical singular limits: 
\bea
x_3\to0: &&3\mbox{ soft}\;,\;
x_3\to1: \, 4\mbox{ soft}\;,\nn\\
x_1\to1: &&3\parallel 1\;,\;
x_1\to0: \, 3\parallel 2\;,\;
x_2\to1: \, 3\parallel 4\;.\nn
\eea
We observe that $\tilde{s}_{14}$ has a line singularity at $x_4=0$, $x_1=x_2$. 
We can decouple the problem at $x_4=0$ from the line singularity $x_1=x_2$
by the transformation~\cite{Anastasiou:2003gr} 
\be
x_4=\frac{t^-\,(1-z_4)}{t^-+z_4\,(t^+ - t^-)}\; \;\Rightarrow \tilde{s}_{14}=\frac{t^+t^-}{t^-+z_4\,(t^+ - t^-)}\;,
\ee
leading to 
\bea
&&\int_0^1 dx_4\,[x_4\,(1-x_4)]^{\frac{D-5}{2}}\,(\tilde{s}_{14})^{-1}=
\int_0^1 dz_4\,[z_4\,(1-z_4)\,t^+t^-]^{\frac{D-5}{2}}\,\left[t^-+z_4\,(t^+ - t^-) \right]^{4-D}\nn\\
&=&\int_0^1 dz_4\,[z_4\,(1-z_4)]^{\frac{D-5}{2}}\,\Big\vert x_1(1-x_2)- x_2(1-x_1)\Big\vert^{D-5}\nn\\
&&\left[(\sqrt{x_1(1-x_2)}-\sqrt{x_2(1-x_1)})^2 +4\,z_4\,\sqrt{x_1(1-x_1)x_2(1-x_2)} \right]^{4-D}\nn\;.
\eea
Now we split the $x_2$-integration range at $x_2=x_1$ and remap to the unit cube:
$$\int_0^1 dx_2 f(x_1,x_2)=\underbrace{\int_0^{x_1} dx_2 f(x_1,x_2)}_{(1)}+
\underbrace{\int_{x_1}^1 dx_2 f(x_1,x_2)}_{(2)}$$
where we substitute $x_2= x_1\,z_2$ in (1) and $x_2 = x_1+(1-x_1)\,z_2$ in (2).
Using the fact that the 
contribution from region (2) equals the first one if we transform $z_2\to1-z_2$ and $x_1\to1-x_1$
and 
combining with the original phase space given in eq.\,(\ref{appendix:psa}), we obtain
\bea
&&\int d\Phi_{3}\,\frac{1}{s_{14}}=
\frac{1}{(2\pi)^{2D-3}}d\Omega_{D-3}d\Omega_{D-4}\,s^{D-4} 2^{D-7}\beta^{2D-6}\label{appendix:pss14}\\
&& \int_0^1 dx_1 dz_2 dx_3 dz_4 [z_4\,(1-z_4)]^{\frac{D-5}{2}}\,x_3^{D-3} \,(1-x_3)^{D-4}
\left[1-\beta\,x_3\,(1-x_1+x_1z_2)\right]^{3-D}\nn\\
&& x_1^{D-4}\,z_2^{D-5}\,\left[(1-x_1)(1-z_2)(1-x_1+x_1z_2)\right]^{\frac{D-4}{2}}\nn\\
&& \left[ (\sqrt{(1-x_1)(1-z_2)}-\sqrt{1-x_1+x_1z_2})^2 +4\,z_4\,\sqrt{(1-x_1)(1-z_2)(1-x_1+x_1z_2)}\right]^{4-D}\;.\nn
\eea

\subsection{Troubleshooting}

Below we give possible reasons and solutions for problems which may arise 
during use of the program.
\begin{itemize}
\item The function ${\cal F}$ is zero:\\
verify the on-shell conditions {\tt onshell=\{\ldots\}} in the file 
{\tt mytemplate.m} where you defined  the integrand.
By default, the external legs have been set to be light-like ($p_i^2=ssp[i]=0$). 
If you calculate a massless two-point integral or a one-scale 3-point integral, 
at least one scale must be different from zero (e.g. set $ssp[1]=-1$ for a two-point 
function with external momentum $p_1$, which amounts to factoring out the overall scale).

Remember that the program by default replaces $p_i^2$ by $ssp[i]$, $(p_i+p_j)^2$ by $sp[i,j]$. 
(This is done in src/deco/calcFU.m). If symbols different from $p_i$ are used for the 
external momenta, the user has to define their numerical values in 
his template file {\tt mytemplate.m}
in the list {\tt onshell}. 

Example: for external vectors called $p$, $q$, 
define numerical values for the invariants formed by $p$ and $q$, 
e.g. {\it onshell}=$\{p^2\to -1,  q^2\to 0, p*q \to -0.5 \}$.
Alternatively, you can map to the predefined names for the invariants, 
e.g. {\it onshell}=$\{p^2\to ssp[1],  q^2\to ssp[2], p*q \to (sp[1,2]-ssp[1]-ssp[2])/2 \}$.
This latter solution allows you to leave the invariants symbolic and specify 
numerical values only at the numerical integration stage, by assigning the corresponding 
numerical values in {\tt param.input}. 

The user can check if the functions  ${\cal F}$, ${\cal U}$ and numerator 
look as expected by looking at the file {\tt FUN.m} in the {\tt integralname/} subdirectory.

\item The numerical integration takes very long:\\
apart from the fact that this is to be expected for complicated integrands, other reasons could be
\begin{itemize}
\item the integrand still contains undefined symbols at the numerical integration stage
because the numerical values for the constants have not been properly defined
(e.g. values for which ${\cal F}$ is not of definite sign, respectively the  general function
develops a singularity within the integration range).
Things to do are: check the function ${\cal F}$ in the file {\tt FUN.m} in the {\tt integralname/} subdirectory
(in the loop case); check the log files of the numerical integration in the subdiretories \\
{\tt integralname/polestructure/epstothe[i]}, where ``polestructure" is of the form e.g. {\tt 2l0h0}, 
denoting 2 logarithmic poles and 0 linear, 0 higher poles.
\item you chose a very large number of Monte Carlo points and/or a very large number of 
iterations  in the input file
\item for functions defined in {\tt SecDec/general}: verify if there is a singularity 
for $x_i\to 1$ rather than only for $x_i\to 0$ and if yes, split this variable at 1/2 by adding its label to the 
{\tt splitlist}.
\end{itemize}

\item the results do not appear in an editor window:\\
either you did not specify an editor in {\tt param.input} (last entry) or 
your system is unable to open the editor window. 
In this case just look at the result file located in the {\tt integralname} subdirectory 
(where {\tt integralname} is the name for the calculated integral or graph, 
specified by you  in {\tt param.input}, third item). The result file is called {\tt integralname\_[point]full.res}.

\item you get the message ``path for Mathematica not automatically found":\\
Insert the path to Mathematica on your system manually for the variable {\tt \$mathpath}
in the file {\tt perlsrc/mathlaunch.pl}.

\end{itemize}



\begin{thebibliography}{99}

\bibitem{Hepp:1966eg}
  K.~Hepp,
  Commun.\ Math.\ Phys.\  {\bf 2} (1966) 301.

\bibitem{Speer:1977uf}
  E.~R.~Speer,
  Annales Poincare Phys.\ Theor.\  {\bf 26} (1977) 87.

\bibitem{Roth:1996pd}
  M.~Roth and A.~Denner,
  Nucl.\ Phys.\  B {\bf 479} (1996) 495
  [arXiv:hep-ph/9605420].

\bibitem{Denner:2004iz}
  A.~Denner and S.~Pozzorini,
  Nucl.\ Phys.\  B {\bf 717} (2005) 48
  [arXiv:hep-ph/0408068].

\bibitem{Binoth:2000ps}
  T.~Binoth and G.~Heinrich,
  Nucl.\ Phys.\  B {\bf 585} (2000) 741
  [arXiv:hep-ph/0004013].

\bibitem{Binoth:2003ak}
  T.~Binoth and G.~Heinrich,
  Nucl.\ Phys.\  B {\bf 680} (2004) 375
  [arXiv:hep-ph/0305234].

\bibitem{Heinrich:2004iq}
  G.~Heinrich and V.~A.~Smirnov,
  Phys.\ Lett.\  B {\bf 598} (2004) 55
  [arXiv:hep-ph/0406053].

\bibitem{Gehrmann:2006wg}
  T.~Gehrmann, G.~Heinrich, T.~Huber and C.~Studerus,
  Phys.\ Lett.\  B {\bf 640} (2006) 252
  [arXiv:hep-ph/0607185].

\bibitem{Heinrich:2007at}
  G.~Heinrich, T.~Huber and D.~Maitre,
  Phys.\ Lett.\  B {\bf 662} (2008) 344
  [arXiv:0711.3590 [hep-ph]].

\bibitem{Heinrich:2009be}
  G.~Heinrich, T.~Huber, D.~A.~Kosower and V.~A.~Smirnov,
  Phys.\ Lett.\  B {\bf 678} (2009) 359
  [arXiv:0902.3512 [hep-ph]].

\bibitem{Baikov:2009bg}
  P.~A.~Baikov, K.~G.~Chetyrkin, A.~V.~Smirnov, V.~A.~Smirnov and M.~Steinhauser,
  Phys.\ Rev.\ Lett.\  {\bf 102} (2009) 212002
  [arXiv:0902.3519 [hep-ph]].

\bibitem{Czakon:2006pa}
  M.~Czakon, J.~Gluza and T.~Riemann,
  Nucl.\ Phys.\  B {\bf 751} (2006) 1
  [arXiv:hep-ph/0604101].

\bibitem{Boughezal:2006xk}
  R.~Boughezal and M.~Czakon,
  Nucl.\ Phys.\  B {\bf 755} (2006) 221
  [arXiv:hep-ph/0606232].

\bibitem{Asatrian:2006ph}
  H.~M.~Asatrian, A.~Hovhannisyan, V.~Poghosyan, T.~Ewerth, C.~Greub and T.~Hurth,
  Nucl.\ Phys.\  B {\bf 749} (2006) 325
  [arXiv:hep-ph/0605009].

\bibitem{Smirnov:2010hd}
  A.~V.~Smirnov and M.~Tentyukov,
  Nucl.\ Phys.\  B {\bf 837} (2010) 40
  [arXiv:1004.1149 [hep-ph]].

\bibitem{Baikov:2010hf}
  P.~A.~Baikov and K.~G.~Chetyrkin,
  Nucl.\ Phys.\  B {\bf 837} (2010) 186
  [arXiv:1004.1153 [hep-ph]].

\bibitem{Lee:2010cga}
  R.~N.~Lee, A.~V.~Smirnov and V.~A.~Smirnov,
  JHEP {\bf 1004} (2010) 020
  [arXiv:1001.2887 [hep-ph]].

\bibitem{Lee:2010ug}
  R.~N.~Lee, A.~V.~Smirnov and V.~A.~Smirnov,
  arXiv:1005.0362 [hep-ph].

\bibitem{Lee:2010ik}
  R.~N.~Lee and V.~A.~Smirnov,
  arXiv:1010.1334 [hep-ph].

\bibitem{Gehrmann:2010ue}
  T.~Gehrmann, E.~W.~N.~Glover, T.~Huber, N.~Ikizlerli and C.~Studerus,
  JHEP {\bf 1006} (2010) 094
  [arXiv:1004.3653 [hep-ph]].

\bibitem{Gehrmann:2010tu}
  T.~Gehrmann, E.~W.~N.~Glover, T.~Huber, N.~Ikizlerli and C.~Studerus,
  arXiv:1010.4478 [hep-ph].

\bibitem{Smirnov:2008py}
  A.~V.~Smirnov and M.~N.~Tentyukov,
  Comput.\ Phys.\ Commun.\  {\bf 180} (2009) 735
  [arXiv:0807.4129 [hep-ph]].

\bibitem{Smirnov:2009pb}
  A.~V.~Smirnov, V.~A.~Smirnov and M.~Tentyukov,
  arXiv:0912.0158 [hep-ph].

\bibitem{Heinrich:2008si}
  G.~Heinrich,
  Int.\ J.\ Mod.\ Phys.\  A {\bf 23} (2008) 1457
  [arXiv:0803.4177 [hep-ph]].

\bibitem{Ferroglia:2002mz}
  A.~Ferroglia, M.~Passera, G.~Passarino and S.~Uccirati,
  Nucl.\ Phys.\  B {\bf 650} (2003) 162
  [arXiv:hep-ph/0209219].

\bibitem{Binoth:2002xh}
  T.~Binoth, G.~Heinrich and N.~Kauer,
  Nucl.\ Phys.\  B {\bf 654} (2003) 277
  [arXiv:hep-ph/0210023].

\bibitem{Lazopoulos:2008de}
  A.~Lazopoulos, T.~McElmurry, K.~Melnikov and F.~Petriello,
  Phys.\ Lett.\  B {\bf 666} (2008) 62
  [arXiv:0804.2220 [hep-ph]].

\bibitem{Lazopoulos:2007ix}
  A.~Lazopoulos, K.~Melnikov and F.~Petriello,
  Phys.\ Rev.\  D {\bf 76} (2007) 014001
  [arXiv:hep-ph/0703273].

\bibitem{Anastasiou:2007qb}
  C.~Anastasiou, S.~Beerli and A.~Daleo,
  JHEP {\bf 0705} (2007) 071
  [arXiv:hep-ph/0703282].

\bibitem{Anastasiou:2008rm}
  C.~Anastasiou, S.~Beerli and A.~Daleo,
  Phys.\ Rev.\ Lett.\  {\bf 100} (2008) 241806
  [arXiv:0803.3065 [hep-ph]].

\bibitem{Soper:1998ye}
  D.~E.~Soper,
  Phys.\ Rev.\ Lett.\  {\bf 81} (1998) 2638
  [arXiv:hep-ph/9804454].

\bibitem{Soper:1999xk}
  D.~E.~Soper,
  Phys.\ Rev.\  D {\bf 62} (2000) 014009
  [arXiv:hep-ph/9910292].

\bibitem{Binoth:2005ff}
  T.~Binoth, J.~P.~Guillet, G.~Heinrich, E.~Pilon and C.~Schubert,
  JHEP {\bf 0510} (2005) 015
  [arXiv:hep-ph/0504267].

\bibitem{Nagy:2006xy}
  Z.~Nagy and D.~E.~Soper,
  Phys.\ Rev.\  D {\bf 74} (2006) 093006
  [arXiv:hep-ph/0610028].

\bibitem{Gong:2008ww}
  W.~Gong, Z.~Nagy and D.~E.~Soper,
  Phys.\ Rev.\  D {\bf 79} (2009) 033005
  [arXiv:0812.3686 [hep-ph]].

\bibitem{Heinrich:2002rc}
  G.~Heinrich,
  Nucl.\ Phys.\ Proc.\ Suppl.\  {\bf 116} (2003) 368
  [arXiv:hep-ph/0211144].

\bibitem{GehrmannDeRidder:2003bm}
  A.~Gehrmann-De Ridder, T.~Gehrmann and G.~Heinrich,
  Nucl.\ Phys.\  B {\bf 682} (2004) 265
  [arXiv:hep-ph/0311276].

\bibitem{Anastasiou:2003gr}
  C.~Anastasiou, K.~Melnikov and F.~Petriello,
  Phys.\ Rev.\  D {\bf 69} (2004) 076010
  [arXiv:hep-ph/0311311].

\bibitem{Binoth:2004jv}
  T.~Binoth and G.~Heinrich,
  Nucl.\ Phys.\  B {\bf 693} (2004) 134
  [arXiv:hep-ph/0402265].

\bibitem{Anastasiou:2004qd}
  C.~Anastasiou, K.~Melnikov and F.~Petriello,
  Phys.\ Rev.\ Lett.\  {\bf 93} (2004) 032002
  [arXiv:hep-ph/0402280].

\bibitem{Anastasiou:2004xq}
  C.~Anastasiou, K.~Melnikov and F.~Petriello,
  Phys.\ Rev.\ Lett.\  {\bf 93} (2004) 262002
  [arXiv:hep-ph/0409088].

\bibitem{Anastasiou:2005qj}
  C.~Anastasiou, K.~Melnikov and F.~Petriello,
  Nucl.\ Phys.\  B {\bf 724} (2005) 197
  [arXiv:hep-ph/0501130].

\bibitem{Anastasiou:2005pn}
  C.~Anastasiou, K.~Melnikov and F.~Petriello,
  JHEP {\bf 0709} (2007) 014
  [arXiv:hep-ph/0505069].

\bibitem{Anastasiou:2007mz}
  C.~Anastasiou, G.~Dissertori and F.~Stockli,
  JHEP {\bf 0709} (2007) 018
  [arXiv:0707.2373 [hep-ph]].

\bibitem{Melnikov:2006di}
  K.~Melnikov and F.~Petriello,
  Phys.\ Rev.\ Lett.\  {\bf 96} (2006) 231803
  [arXiv:hep-ph/0603182].

\bibitem{Melnikov:2006kv}
  K.~Melnikov and F.~Petriello,
  Phys.\ Rev.\  D {\bf 74} (2006) 114017
  [arXiv:hep-ph/0609070].

\bibitem{Melnikov:2008qs}
  K.~Melnikov,
  Phys.\ Lett.\  B {\bf 666} (2008) 336
  [arXiv:0803.0951 [hep-ph]].

\bibitem{Biswas:2009rb}
  S.~Biswas and K.~Melnikov,
  JHEP {\bf 1002} (2010) 089
  [arXiv:0911.4142 [hep-ph]].

\bibitem{Gavin:2010az}
  R.~Gavin, Y.~Li, F.~Petriello and S.~Quackenbush,
  arXiv:1011.3540 [hep-ph].

\bibitem{Frixione:1995ms}
  S.~Frixione, Z.~Kunszt and A.~Signer,
  Nucl.\ Phys.\  B {\bf 467} (1996) 399
  [arXiv:hep-ph/9512328].

\bibitem{Czakon:2010td}
  M.~Czakon,
  Phys.\ Lett.\  B {\bf 693} (2010) 259
  [arXiv:1005.0274 [hep-ph]].

\bibitem{Anastasiou:2010pw}
  C.~Anastasiou, F.~Herzog and A.~Lazopoulos,
  arXiv:1011.4867 [hep-ph].

\bibitem{Smirnov:2006ry}
  V.~A.~Smirnov,
  {\it Feynman integral calculus},
Springer, Berlin (2006),  283 p.

\bibitem{Bogner:2007cr}
  C.~Bogner and S.~Weinzierl,
  Comput.\ Phys.\ Commun.\  {\bf 178} (2008) 596
  [arXiv:0709.4092 [hep-ph]].

\bibitem{Hironaka}
H. Hironaka,  Ann. Math. {\bf 79} (1964) 109.

\bibitem{Gluza:2010rn}
  J.~Gluza, K.~Kajda, T.~Riemann and V.~Yundin,
  arXiv:1010.1667 [hep-ph].

\bibitem{Smirnov:2008aw}
  A.~V.~Smirnov and V.~A.~Smirnov,
  JHEP {\bf 0905} (2009) 004
  [arXiv:0812.4700 [hep-ph]].

\bibitem{Ueda:2009xx}
  T.~Ueda and J.~Fujimoto,
  PoS  {\bf ACAT08} (2008) 120
  [arXiv:0902.2656 [hep-ph]].

\bibitem{Vermaseren:2000nd}
  J.~A.~M.~Vermaseren,
  {\it New features of FORM},
  arXiv:math-ph/0010025.

\bibitem{Kaneko:2009qx}
  T.~Kaneko and T.~Ueda,
  Comput.\ Phys.\ Commun.\  {\bf 181} (2010) 1352
  [arXiv:0908.2897 [hep-ph]].

\bibitem{Kaneko:2010kj}
  T.~Kaneko and T.~Ueda,
  arXiv:1004.5490 [hep-ph].

\bibitem{nakanishi}
N. Nakanishi,
\newblock Graph Theory and Feynman Integrals (Gordon and Breach, New York,
  1971).

\bibitem{Tarasov:1996br}
  O.~V.~Tarasov,
  Phys.\ Rev.\  D {\bf 54} (1996) 6479
  [arXiv:hep-th/9606018].

\bibitem{Landau:1959fi}
  L.~D.~Landau,
  Nucl.\ Phys.\  {\bf 13} (1959) 181.

\bibitem{ELOP}
R.J. Eden et~al.,
\newblock The Analytic S-Matrix (Cambridge University Press, 1966).


\bibitem{Tkachov:1997ap}
  F.~V.~Tkachov,
  Int.\ J.\ Mod.\ Phys.\  A {\bf 14} (1999) 683
  [arXiv:hep-ph/9703423].

\bibitem{Gelfand}
I. Gelfand and G. Shilov,
\newblock {Generalized Functions}, Volume~1 (Academic Press, New York, 1964).

\bibitem{Wolfram}
Mathematica, Copyright by Wolfram Research.

\bibitem{Kawabata:1995th}
  S.~Kawabata,
  Comput.\ Phys.\ Commun.\  {\bf 88} (1995) 309.

\bibitem{Hahn:2004fe}
  T.~Hahn,
  Comput.\ Phys.\ Commun.\  {\bf 168} (2005) 78
  [arXiv:hep-ph/0404043].

\bibitem{Format}
M. Sofroniou,
\newblock http://library.wolfram.com/infocenter/MathSource/60/  (2005).

\bibitem{robodoc}
F. Slothouber  et~al.,
 http://www.xs4all.nl/\textasciitilde rfsber/Robo/robodoc.html .

\bibitem{Huber:2005yg}
  T.~Huber and D.~Maitre,
  Comput.\ Phys.\ Commun.\  {\bf 175} (2006) 122
  [arXiv:hep-ph/0507094].

\bibitem{Huber:2007dx}
  T.~Huber and D.~Maitre,
  Comput.\ Phys.\ Commun.\  {\bf 178} (2008) 755
  [arXiv:0708.2443 [hep-ph]].

\bibitem{Reiter:2009ts}
  T.~Reiter,
  Comput.\ Phys.\ Commun.\  {\bf 181} (2010) 1301
  [arXiv:0907.3714 [hep-ph]].
\end{thebibliography}
\end{document}